%% file: main.tex
\documentclass[10pt,journal,compsoc]{IEEEtran}

\ifCLASSOPTIONcompsoc
  \usepackage[nocompress]{cite}
\else
  \usepackage{cite}
\fi
\ifCLASSINFOpdf
\else
\fi

\usepackage[pdftex]{graphicx}
\usepackage{caption}
\usepackage{subcaption}
\usepackage{cite}

\usepackage{multirow}
\usepackage{booktabs}
\usepackage{tablefootnote}

\usepackage{url}

\usepackage{breakurl}
\usepackage[breaklinks]{hyperref}

\usepackage{xcolor}
\usepackage{amsmath}

\usepackage{enumitem}

\setenumerate[1]{itemsep=0.5pt,partopsep=0pt,parsep=\parskip, topsep=2pt, leftmargin=12pt,}

\newcommand{\toolName}{TopSort}

\begin{document}
%
\title{\huge{\toolName: A High-Performance Two-Phase Sorting Accelerator Optimized on HBM-based FPGAs}}
%
%
%
%

\author{Weikang~Qiao,~\IEEEmembership{Student~Member,~IEEE,}
        Licheng~Guo,
        Zhenman~Fang,~\IEEEmembership{Member,~IEEE,}
        Mau-Chung~Frank~Chang,~\IEEEmembership{Life~Fellow,~IEEE}
        and~Jason Cong,~\IEEEmembership{Fellow,~IEEE}
\IEEEcompsocitemizethanks{\IEEEcompsocthanksitem Weikang Qiao and Mau-Chung Frank Chang are with the Department of Electrical and Computer Engineering, University of California, Los Angeles, CA, 90025.\protect\\
E-mail: wkqiao2015@ucla.edu, mfchang@ee.ucla.edu
\IEEEcompsocthanksitem Licheng Guo and Jason Cong are with the Department of Computer Science, University of California, Los Angeles, CA, 90025.\protect\\
E-mail: \{lcguo,cong\}@cs.ucla.edu
\IEEEcompsocthanksitem Zhenman Fang is with the School of Engineering Science, Simon Fraser University, Burnaby, BC V5A 1S6, Canada.\protect\\
E-mail: zhenman@sfu.ca
}
}
\IEEEtitleabstractindextext{%
\input{0-abstract}

}

\maketitle

\IEEEdisplaynontitleabstractindextext

%
\IEEEpeerreviewmaketitle

\input{1-introduction}
\input{2-motivation}
\input{3-methodology}
\input{4-hbm_optimize}
\input{5-floorplan}

\input{6-experiment}

\input{7-related_work}
\input{8-conclusion}
\input{9-acknowledgement}

\ifCLASSOPTIONcaptionsoff
  \newpage
\fi

\bibliographystyle{IEEEtran}
\bibliography{references.bib}

\input{biography}

\end{document}

%% file: 0-abstract.tex
\begin{abstract}

The emergence of high-bandwidth memory (HBM) brings new opportunities to boost the performance of sorting acceleration on FPGAs, which was conventionally bounded by the available off-chip memory bandwidth. However, it is nontrivial for designers to fully utilize this immense bandwidth. First, the existing sorter designs cannot be directly scaled at the increasing rate of available off-chip bandwidth, as the required on-chip resource usage grows at a much faster rate and would bound the sorting performance in turn. Second, designers need an in-depth understanding of HBM’s characteristics to effectively utilize the HBM bandwidth. To tackle these challenges, we present TopSort, a novel two-phase sorting solution optimized for HBM-based FPGAs. In the first phase, 16 merge trees work in parallel to fully utilize 32 HBM channels’ bandwidth. In the second phase, TopSort reuses the logic from phase one to form a wider merge tree to merge the partially sorted results from phase one. TopSort also adopts HBM-specific optimizations to reduce resource overhead and improve bandwidth utilization. TopSort can sort up to 4 GB data using all 32 HBM channels, with an overall sorting performance of 15.6 GB/s. TopSort is 6.7$\times$ and 2.2$\times$ faster than state-of-the-art CPU and FPGA sorters.

\end{abstract}

%% file: 1-introduction.tex
\section{Introduction} \label{sec:intro}
Sorting is one of the fundamental computation kernels in many big data applications and there have been continuous efforts on designing high-performance sorting accelerators on FPGAs~\cite{fpga'11, fpga'14, fpga'16, fccm'16,  fccm'17-tera, fpt'18, isca'20, fpl'20, fccm'21, fpga'15, fccm'17-tokyo, fccm'18, fpt'18-merger, mcsoc'19, fpga'20}. Most of these sorting accelerators are based on the multi-way merge tree sort algorithm~\cite{fpga'11, fpga'14, fpga'16, fccm'16, fccm'17-tera, fpt'18, isca'20, fpl'20, fccm'21}, due to its massive data parallelism and regular memory access patterns.
Before the maturing of the high-bandwidth memory (HBM) technology, these sorters were usually implemented on DRAM-based FPGAs and bounded by the off-chip memory bandwidth~\cite{isca'20, fpl'20}.
For example, Bonsai~\cite{isca'20} presented the state-of-the-art merge tree based sorter on the AWS F1 datacenter FPGA and achieved an overall sorting performance of 7.1 GB/s with a merge tree's throughput of 32 GB/s;\footnote{The definitions of a merge tree's throughput and the overall sorting performance are presented in Section~\ref{sec:background} and Section~\ref{subsec:analysis_single_tree}.} it could sufficiently scale up the number of parallel merge units until the memory bandwidth becomes the bottleneck. 




The emergence of HBM-based FPGAs has brought the potential to further boost the performance of the sorting accelerators by offering a much higher off-chip memory bandwidth. For example, the recent HBM-based datacenter FPGAs report a peak memory bandwidth of about 420 GB/s, 
which is roughly 6$\times$ of that on a similar DRAM-based datacenter FPGA~\cite{fpga'21-ubench,fpga'21-hbmconnect}. A natural question is: if we simply port and scale the state-of-the-art merge tree designs onto an HBM-based FPGA, can we get 6$\times$ higher merge tree throughput? Unfortunately, the answer is no due to two nontrivial challenges.

On the one hand, with the tremendous off-chip bandwidth increase, the bottleneck of sorting acceleration would shift from off-chip memory to the available on-chip resources. As will be analyzed in Section~\ref{subsec:analysis_single_tree}, for a merge tree-based sorter that can output $p$ elements per cycle, the required off-chip bandwidth increases linearly with $p$, while the required on-chip resources increase much faster at the rate of $\theta(plog^3(p))$. Unfortunately, for an HBM-based FPGA (e.g., Xilinx Alveo U280 FPGA~\cite{u280}) that provides roughly 6$\times$ more bandwidth than a similar DRAM-based FPGA (e.g., AWS F1 FPGA), it only provides merely 1.2$\times$ more on-chip resources. Indeed, our analysis shows that a single merge tree accelerator can not be scaled to use more than $1/4$ of the available HBM bandwidth.



On the other hand, it is nontrivial to fully utilize the HBM bandwidth in an accelerator design.
Typically, the HBM is composed of 32 small channels. The accelerator has to rely on multiple memory controllers to access those HBM channels in parallel to maximize the memory bandwidth, at the expense of on-chip resources. Moreover, there could easily be contention between multiple channel accesses due to HBM's internal channel switching, which would degrade the effective bandwidth. Therefore, it requires delicate memory access control to improve the bandwidth utilization and reduce resource overhead. Finally, the HBM stacks are physically connected to a datacenter FPGA's bottom die only, making it difficult to spread the resource utilization across multiple FPGA dies to achieve desirable timing closure. To the best of our knowledge, none of the published HBM-based accelerator designs~\cite{iccad'21, arxiv'21-serpens, fpt'21, fpga'22-spmv,fpga'22-sssp} is able to fully utilize the entire bandwidth of the 32 HBM channels. 

In this work, we present \toolName, a high-performance \underline{t}w\underline{o}-\underline{p}hase \underline{sort}ing accelerator specialized for HBM-based FPGAs. \toolName~avoids the excessive resource consumption of directly scaling a single giant sorter's throughput by novelly splitting the complete sorting process into two separate merge phases with smaller sorters and reusing the resources between the two phases. In the first phase, \toolName~employs 16 parallel small merge tree kernels, each of which sorts a portion of the input sequence with two HBM channels (one for reading unsorted inputs and the other for writing sorted outputs). In this phase, the effective merge tree throughput is equal to the bandwidth of 16 HBM channels. 
In the second phase, \toolName~merges the sorted results from all HBM channels into one final sorted sequence. To reduce the resource consumption, this phase reuses 4 merge tree kernels from phase one to form a wider merge tree with a 4$\times$ higher throughput, since merge trees with different throughput share a similar operation pattern. 

To improve the effective HBM bandwidth utilization, we profile the corresponding HBM channel characteristics and carefully optimize the merge tree's memory access pattern for each phase, including optimizing its data layout and burst access, as well as reducing the usage of HBM controllers.
Besides, the novel merge tree reuse architecture allows us to easily improve the design frequency through coarse-grained floorplanning of each separate merge tree kernel. We floorplan \toolName~by evenly distributing the design across the entire FPGA logic regions, based on an efficient resource model that considers the accelerator's design complexity, the available resources of each FPGA die, the limitation of cross-die signals, and the overhead of HBM controllers.

When implemented on the Xilinx Alveo U280 FPGA, \toolName~runs at 214 MHz and achieves an overall performance of 15.6 GB/s, which is 6.7$\times$ and 2.2$\times$ faster than state-of-the-art CPU and FPGA sorters~\cite{vldb15-radix,isca'20}. \toolName~can sort up to 4 GB data at a time, which is half of the total HBM capacity.
Although this work includes a number of device-specific optimization for the HBM-based U280 FPGAs from Xilinx, the general two-phase reused-based merge sort architecture is applicable to all HBM-based FPGAs across different vendors. 

We summarize the contributions of \toolName~as below:

\begin{enumerate}

\item A novel two-phase merge tree based sorter optimized on HBM-based FPGAs, which fully utilizes the HBM bandwidth and alleviates the on-chip resource bottleneck by intelligently reusing multiple merge trees between two phases. 

\item Techniques and insights for HBM-specific optimizations, including the data layout, burst access, and HBM controller optimizations, as well as the floorplanning strategy.

\item Experimental results that demonstrate the superior 15.6 GB/s sorting performance and show~\toolName~is 6.7$\times$ and 2.2$\times$ faster than state-of-the-art CPU and FPGA sorters.

\end{enumerate}





%% file: 2-motivation.tex
\section{Background Review} \label{sec:background}
The merge tree sorting algorithm is favored for FPGA-based sorters due to its massive data parallelism, less control overhead and regular memory access patterns~\cite{fpga'11, fpga'14, fpga'16, fccm'16, fccm'17-tera, fpt'18, isca'20, fpl'20, fccm'21}. In this section, we first introduce the HBM-based FPGAs. Then we give an overview of the hardware merge units and the existing DRAM-based merge tree sorting accelerators. 

\subsection{HBM-Based FPGAs}

\begin{figure}[!t]
    \centering
    \includegraphics[width=0.8\columnwidth]{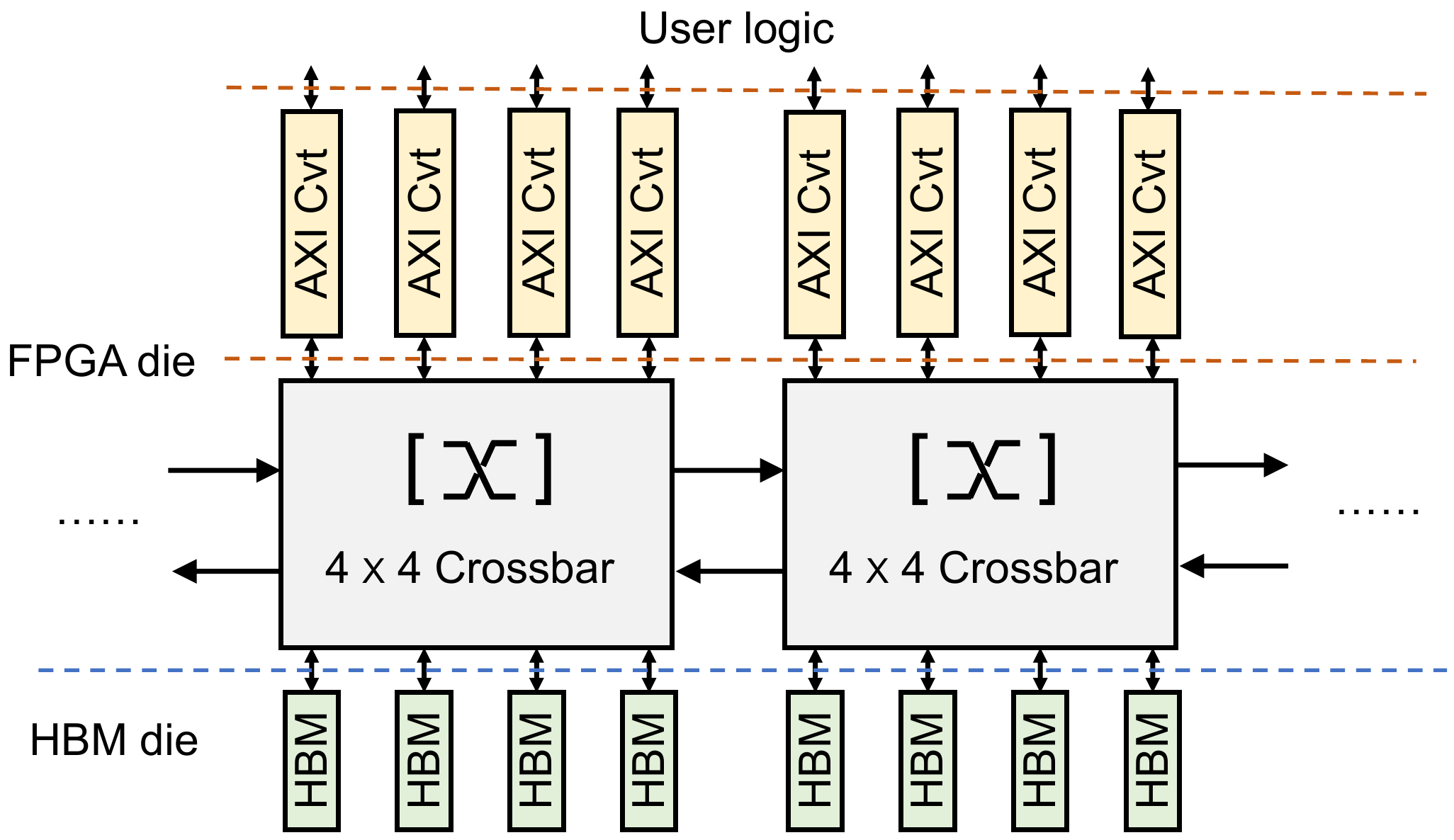}
    \caption{Connections from user logic to HBM channels. AXI Cvt is short for AXI Rate Converter. Lateral connections between nearby crossbars enable AXI to access HBM channels that are not located in the same group.}
    \label{fig:hbm_axi}
\end{figure}

HBM achieves higher bandwidth than DDR4 DRAMs by stacking multiple small DRAM dies together and is one of the most promising candidates enabling memory-centric designs~\cite{pact'15}. Taking Xilinx U280 board as an example, the board is equipped with 2 HBM stacks and each stack contains 16 pseudo channels~\cite{u280}. Figure~\ref{fig:hbm_axi} exhibits the connections between the user logic and the HBM channels. Each HBM channel can be accessed through a 256-bit wide AXI interface running at 450 MHz. The vendor tool will by default implement AXI rate converters in the HBM Memory Subsystem (HMSS) to adapt the original 256-bit AXI interfaces to 512-bit AXI interfaces running at 225 MHz to the user logic. In the case of routing congestion which happens if the on-chip resources are over-utilized or the design is not well pipelined, both the HBM-side AXI frequency and the user-side design frequency will be reduced and the available HBM bandwidth will be degraded.

There is no global crossbar to allow 32 AXI interfaces to access the 32 HBM channels at the same time. Instead, the 32 HBM channels are physically bundled into 8 groups and each group contains 4 adjacent channels joined by a built-in 4$\times$4 crossbar, as is shown in Figure~\ref{fig:hbm_axi}. The crossbar provides full connectivity within the group. Meanwhile, each AXI interface at the user side can still access any HBM channels outside its group. The data will sequentially traverse through each of the lateral connections until it reaches the crossbar connecting to the target channel. 

\subsection{Hardware Merge Unit} \label{subsec: parallel_merger}
\begin{figure}[!t]
 \begin{center}
 \includegraphics[width=3.2in]{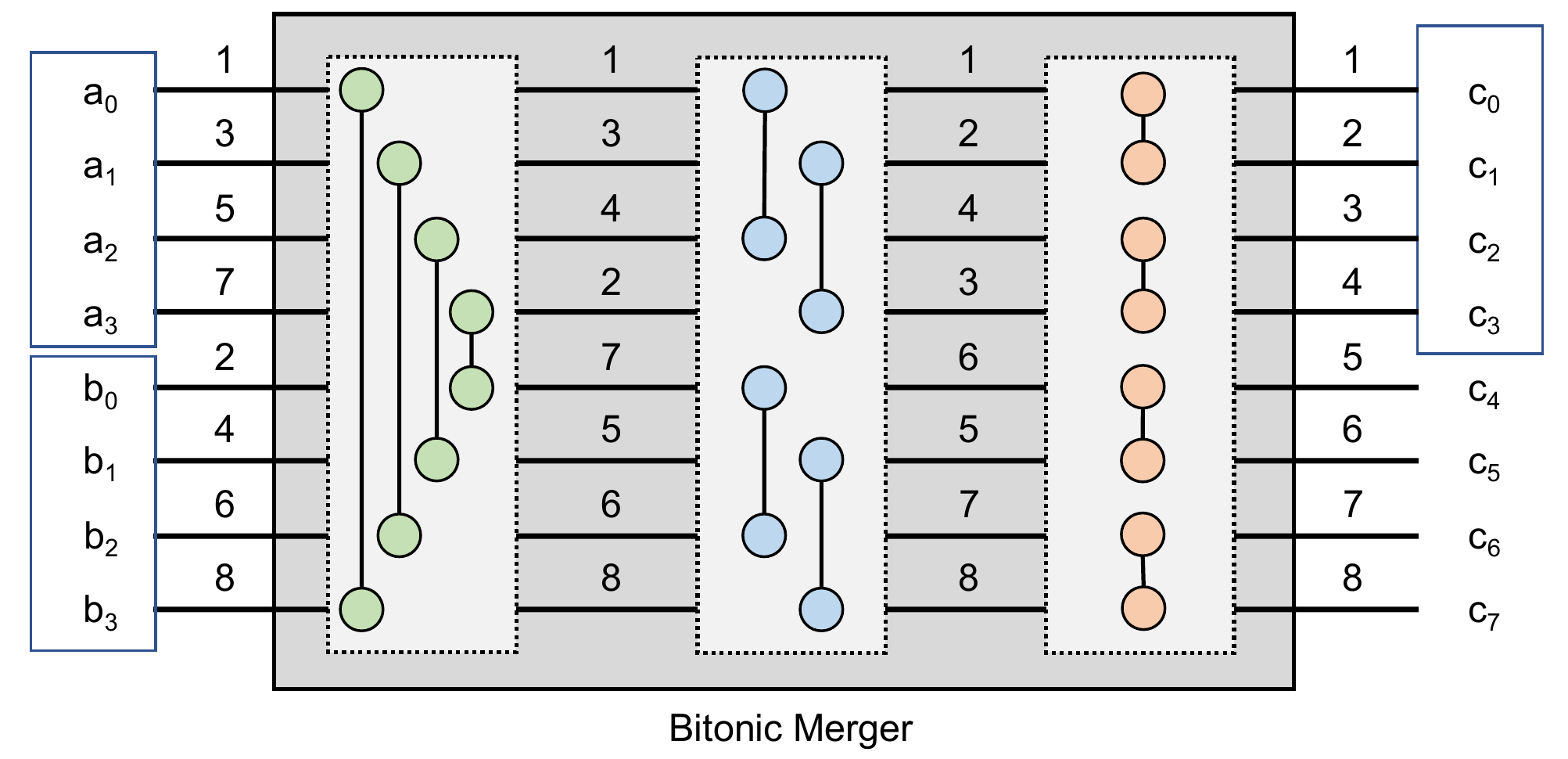}
 \caption{An example of $4$-rate bitonic merge unit: each vertical line that connects two dots is a compare-swap cell and the compare-swap cells in the same box are processed in the same cycle. $c_{0-3}$ will be the outputs after 3 cycles.}
 \label{fig:bitonic}
 \end{center}
 \end{figure}

A hardware merge unit takes two sorted sequences of elements as its inputs and then merges them into one sorted sequence. Specifically, an $E$-rate hardware merge unit takes $E$ input elements from its two $E$-element wide inputs and outputs $E$ sorted elements every cycle. The \textit{compare-swap cell} is the basic building block for hardware merge unit, which compares two elements' values and swaps them into the correct ordering~\cite{fpga'11}. A compare-swap cell usually contains a comparator and a 2-input multiplexer, which is suitable to be implemented using the Look-Up Tables (LUTs) on the FPGAs. Using a pipeline of multiple compare-swap cells, designers can develop the hardware merge unit. The bitonic merge unit shown in Figure~\ref{fig:bitonic} is one of the most widely used hardware merge units.

 \begin{figure}[!t]
 \begin{center}
 \includegraphics[width=3.2in]{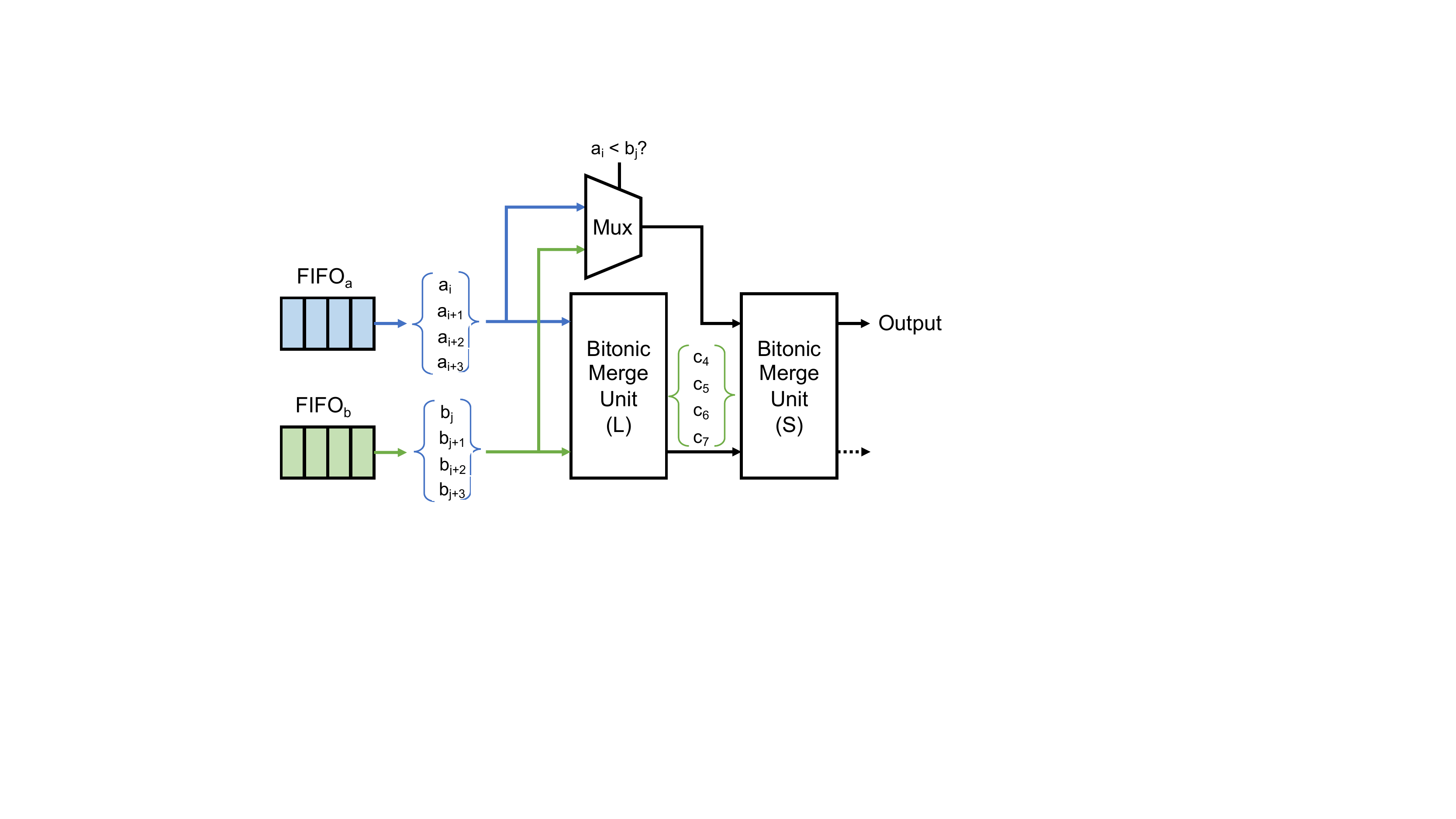}
 \caption{The topology of a $4$-rate MMS merge unit. Registers in the datapath are omitted for simplicity.}
 \label{fig:shms}
 \end{center}
 \end{figure}
 
If the two sorted input sequences are longer than $E$, the $E$-rate merge unit has to be time multiplexed (i.e., executed multiple rounds) with extra control signals to ensure the outputs are in order. For example, in Figure~\ref{fig:bitonic}, $c_{4-7}$ represent the largest four elements of $a_{0-3}$ and $b_{0-3}$. In the next cycle, $c_{4-7}$ need to be sent back to the input side of the same bitonic merge unit and merged with either $a_{4-7}$ or $b_{4-7}$ to create the second $4$-element sorted outputs. The feedback paths from the output side of the bitonic merge unit to its input side mean the merge unit has to wait for extra cycles before it can do the next merge operation. In other words, the initial interval (II) is equal to the number of pipeline stages in the bitonic merge unit.

To improve the merging performance with an II equal to 1, several optimization techniques have been proposed~\cite{fccm'17-tokyo,fpt'18-merger,fccm'18,mcsoc'19}.  \cite{fccm'18} proposes an $E$-rate streaming merge unit called \textit{MMS} that outputs $E$ elements every cycle using two bitonic merge units, as shown in Figure~\ref{fig:shms}. The intuition is that the larger half outputs of the original bitonic merge unit can be first calculated through the bitonic merge unit (L) and are later fed into the bitonic merge unit (s) with delayed inputs from either sequence $a$ or $b$. In this work, we use the same topology of MMS to construct the hardware merge unit, as it is free of feedback paths and thus helps alleviate the design routing congestion.


\subsection{DRAM-based Merge Tree Sorting Accelerator} \label{subsec:existing_tree}

\begin{figure}[!t]
    \centering
    \includegraphics[width=\columnwidth]{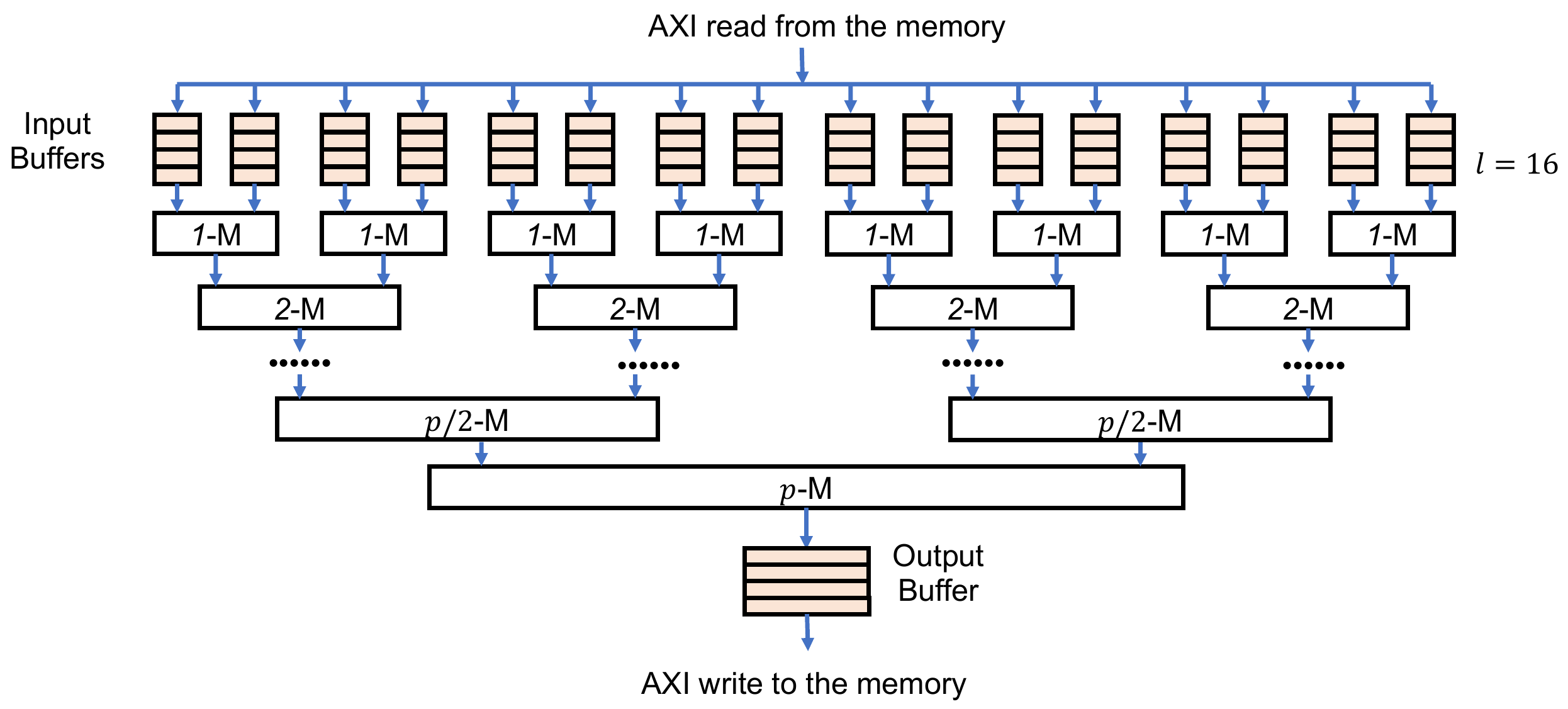}
    \caption{An example of the merge tree ($p=16$, $l=16$), where an $E$-M box denotes an $E$-rate hardware merge unit. Note that $l$ can be larger than $p$, e.g., to make a merge tree ($p=16$, $l=32$), one can have another layer of 32 $1$-M merge units on top of the current leaf layer.}
    \label{fig:single_merge_tree}
\end{figure}

Using a combination of various hardware merge units with different rates, we can build a complete binary tree that consumes $l$ unsorted input sequences concurrently at its leaves and outputs $p$ sorted elements at its root every cycle~\cite{fccm'16,fpt'18,isca'20,fpl'20}. Figure~\ref{fig:single_merge_tree} shows the architecture of a merge tree~\cite{isca'20}. Such a merge tree can be uniquely denoted by ($p$, $l$), where $p$ refers to the \textbf{merge tree's throughput} and $l$ is \textbf{the number of leaves}.

In one sorting \textit{pass}, the input elements are streamed from the off-chip memory into the leaf buffers and then through the merge tree, and the output elements are streamed back into the off-chip memory. Assume there are $N$ unsorted elements initially stored in the off-chip memory, these $N$ elements need to be streamed into the merge tree for multiple passes and after each pass, the size of the partially sorted sequence grows exactly by $l$ times. 
During the first pass, the merge tree reads these $N$ sub sequences (each containing one element) from the off-chip memory, merges them into $N/l$ sorted sub sequences, each containing $l$ sorted elements, and writes them back to the off-chip memory. In the next pass, the $l$-element sorted sub sequences are fed into the merge tree again to get $N/l^2$ sorted sequences of length $l^2$. These steps are repeated until the last pass, where $l$ sorted sequences of length $N/l$ are processed by the merge tree to form a complete $N$-element sorted sequence. One can easily derive that the total number of passes is $\lceil \log_{\ell} N \rceil$.


\section{Scalability Analysis of a Single Merge Tree} \label{subsec:analysis_single_tree}
In this section, we explain why the single merge tree employed on existing DRAM-based FPGAs cannot be directly scaled to efficiently work on HBM-based FPGAs.

The throughput of a single merge tree can be derived using the analytical model from \cite{isca'20}. First, the merge tree with $l$ leaves takes $\lceil \log_{\ell} N \rceil$ passes. Second, the merge tree outputs $p$ elements every cycle to the off-chip memory,
where $p$ is directly reflected by the off-chip memory bandwidth $\beta_{memory}$ allocated to the writing operations. Generally, half of the system bandwidth is allocated for writing and the other half is for reading. We define the \textbf{overall sorting performance} as the number of bytes of unsorted elements divided by the time it takes to get the completely sorted elements. Assuming there are unlimited on-chip resources, we have the overall sorting performance $\beta_{overall}$ in Equation~\ref{eq:bonsai_perf}.
 \begin{equation}\label{eq:bonsai_perf}
     \beta_{overall} = \frac{\beta_{memory}}{\lceil \log_{\ell} N \rceil}  
 \end{equation}

Clearly, the on-chip resources are finite. Next, we analyze the on-chip resource requirement in scaling the merge tree-based sorter. 
The merge tree in Figure~\ref{fig:single_merge_tree} can be viewed as two sub merge trees whose root rates are $p/2$ each, plus a merge unit whose rate is $p$. Since a merge unit is a variation of a parallel sorting network, such as bitonic or even-odd sorting network, and such a $p$-rate merge unit consumes $plog^2(p)$ number of comparison operators~\cite{fccm'17-tokyo, fccm'18}, the number of comparators, $L(p)$, required when scaling $p$ can be summarized in Equation~\ref{eq:tree_scale}.
\begin{equation}\label{eq:tree_scale}
    L(p) = 2L(p/2)+\theta(plog^2(p))  
\end{equation}

Based on this equation, we can derive that the total number of logic elements (i.e., comparators) for the complete binary tree is $\theta(plog^3(p))$~\cite{bentley1980general}.
Considering that migrating from 4 DRAM channels in~\cite{isca'20} to 32 HBM channels increases the available off-chip bandwidth by nearly 8$\times$, it is not possible to directly scale a single merge tree's throughput to catch up with the increase of the HBM bandwidth. 

In fact, we find that the root throughput of a single merge tree can only be scaled to match 4 HBM channels (for write operations) on the Xilinx Alveo U280 board. Since the read operations of the tree also occupies the same amount of HBM bandwidth, the bandwidth utilized in such a merge tree is at most equal to 8 HBM channels. In other words, such a merge tree only uses 25\% of the HBM bandwidth in each pass.

%% file: 3-methodology.tex
\section{\toolName~Methodology \& Architecture} \label{sec:analysis}

In this section, we present the methodology of \toolName. First, we illustrate the idea of the two-phase sorting. Second, we show how \toolName~reuses the logic between two phases to alleviate the resource contention and improve the performance. 

\subsection{Two Phases in \toolName} \label{subsec:analy_2_phase}
Since the required resources grow super-linearly when directly scaling a single merge tree, \toolName~chooses to split the $N$ unsorted elements evenly into $k$ parts and have $k$ smaller merge trees work in parallel. Each merge tree sorts $N/k$ elements and is configured to have a throughput that saturates the bandwidth of a single HBM channel. This method has better scalability as its resource consumption grows linearly with the number of trees $k$. However, the $k$ sorted sequences still need to be merged into the final sorted sequence. This can be done by streaming the $k$ sequences into another merge tree for the final merge. We refer to the process during which $k$ merge trees work in parallel as phase 1, and the subsequent final merge as phase 2.


If each merge tree in phase 1 has $l_{phase1}$ number of leaves and a single HBM channel's bandwidth is $\beta_{channel}$, the average performance in phase 1 is given in Equation~\ref{eq:phase1_perf}. The fraction part is a single merge tree's performance when sorting $N/k$ elements, and there are $k$ such merge trees in parallel.  
\begin{equation} \label{eq:phase1_perf}
    \beta_{phase1} = k \cdot \frac{\beta_{channel}}{\lceil \log_{l_{phase1}} (N/k) \rceil}
\end{equation}

Let $p_{phase2}$ denote the merge tree's throughput in phase 2. As long as the merge tree has $k$ or more leaves, phase 2 will require only \textit{one} pass, e.g., if phase 2's merge tree has $k$ leaves, then each of the $k$ sorted sequences from phase 1 will be fed into one of the corresponding $k$ leaves once. According to the analysis in Section~\ref{subsec:analysis_single_tree}, the merge tree of phase 2 only exploits a portion of the HBM bandwidth due to on-chip resource constraints. Therefore, the performance $\beta_{phase2}$ of phase 2 is proportional to $p_{phase2}$.

The overall sorting performance is in Equation~\ref{eq:cmpl_perf}.
\begin{equation} \label{eq:cmpl_perf}
    \beta_{overall} = \frac{1}{\frac{1}{\beta_{phase1}}+\frac{1}{\beta_{phase2}}}
\end{equation}
This highlights that the resource balance between the parallel merge trees of phase 1 and the final merge tree of phase 2 plays an important role in the overall sorting performance. For example, we may have 16 merge trees in the first phase to fully utilize all 32 HBM channels and achieve the best performance. But if the remaining resources only allow us to build a final merge tree whose throughput is equal to the bandwidth of a single channel, then the overall sorting performance will be limited by the final merge, no matter how much better the memory bandwidth utilization we achieve in phase 1. One may wonder if we can reprogram the FPGA in phase 2 to have a wider tree for final merging, but the reprogramming overhead takes several seconds~\cite{fccm'21} and thus is not practical in this case.

%


\begin{figure}[!t]
    \centering
    \includegraphics[width=0.9\columnwidth]{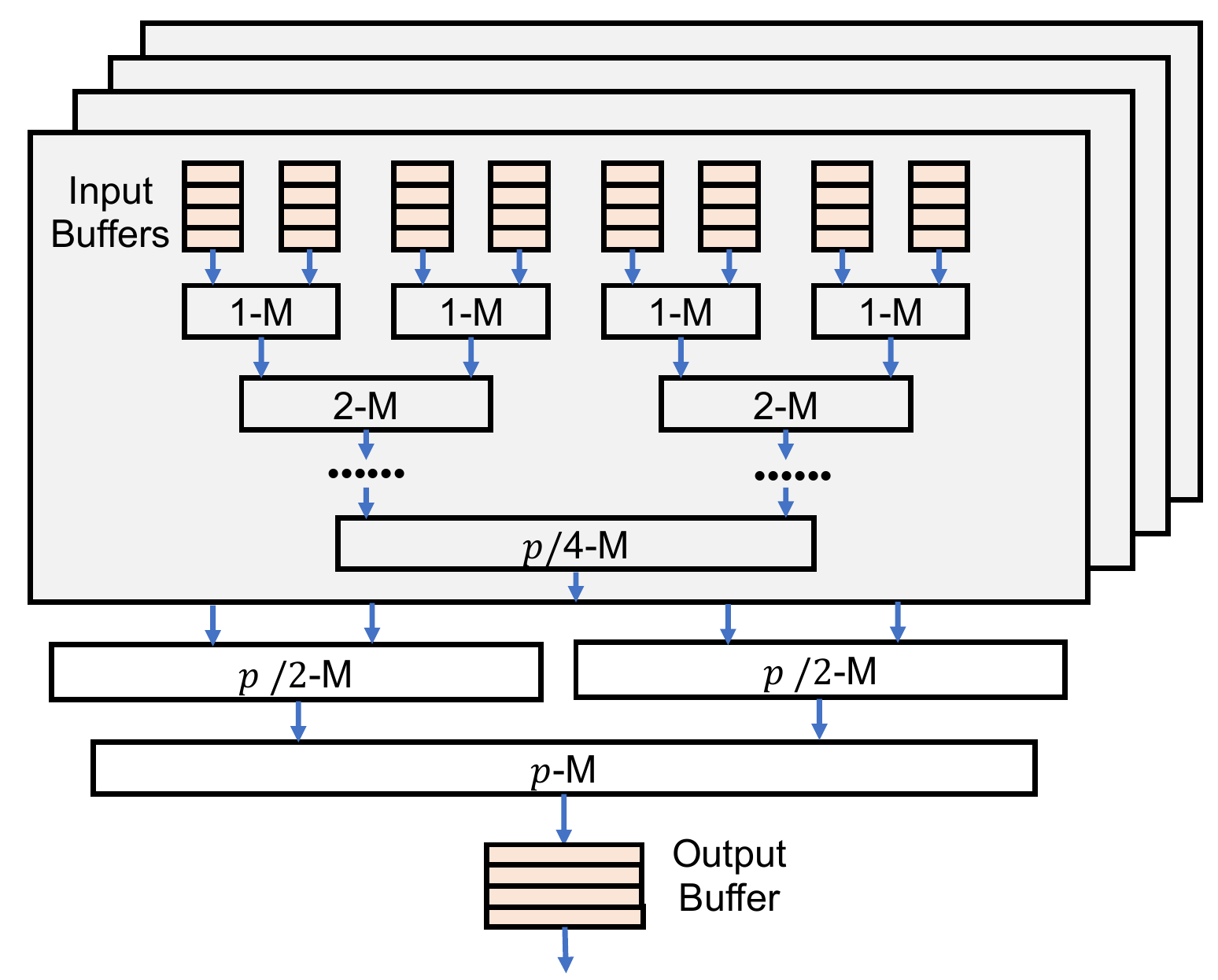}
    \caption{Merge tree reuse: four $p/4$-rate merge trees in phase 1 are reused to build a larger $p$-rate merge trees in phase 2.}
    \label{fig:reuse_merge_tree}
\end{figure}

\subsection{Merge Tree Reuse in \toolName}
To resolve the resource contention between phases 1 \& 2, \toolName~proposes to reuse the merge units in phase 1 to build the final merge tree. This is based on the observation that a merge tree can be decomposed as sub merge trees plus some extra levels of merge units. For example, a merge tree whose throughput is $p$ in Figure~\ref{fig:single_merge_tree} consists of four merge trees of throughput $p/4$, plus two $p/2$-rate merge units and one $p$-rate merge unit, as shown in Figure~\ref{fig:reuse_merge_tree}. In other words, if we want to build a merge tree whose throughput is equal to 4 HBM channels' bandwidth in phase 2, we only need extra resources to build the 3 more merge units. The rest of the resources including the input buffers can be reused from four of the merge trees in phase 1.
With this novel merge tree reuse, conceptually, \toolName~is able to integrate 16 merge trees to saturate all 32 HBM channels' bandwidth in phase 1 and another merge tree whose throughput saturates 4 HBM channels' bandwidth in phase 2 onto one FPGA. 

Please note this merge tree reuse approach shows a general architecture that could be easily scaled with different hardware configurations. Given the available off-chip bandwidth (e.g., the number of HBM channels) and on-chip resources (e.g., available BRAMs and LUTs), we can easily adapt (either scale up or down) the number of the parallel merge trees in phase 1, the size (i.e., number of tree leaves l) of each merge tree in phase 1 and the final merge tree in phase 2.

\subsection{Architecture to Support Logic Reuse} \label{subsec:arch}
\begin{figure}[!t]
    \centering
    \includegraphics[width=\columnwidth]{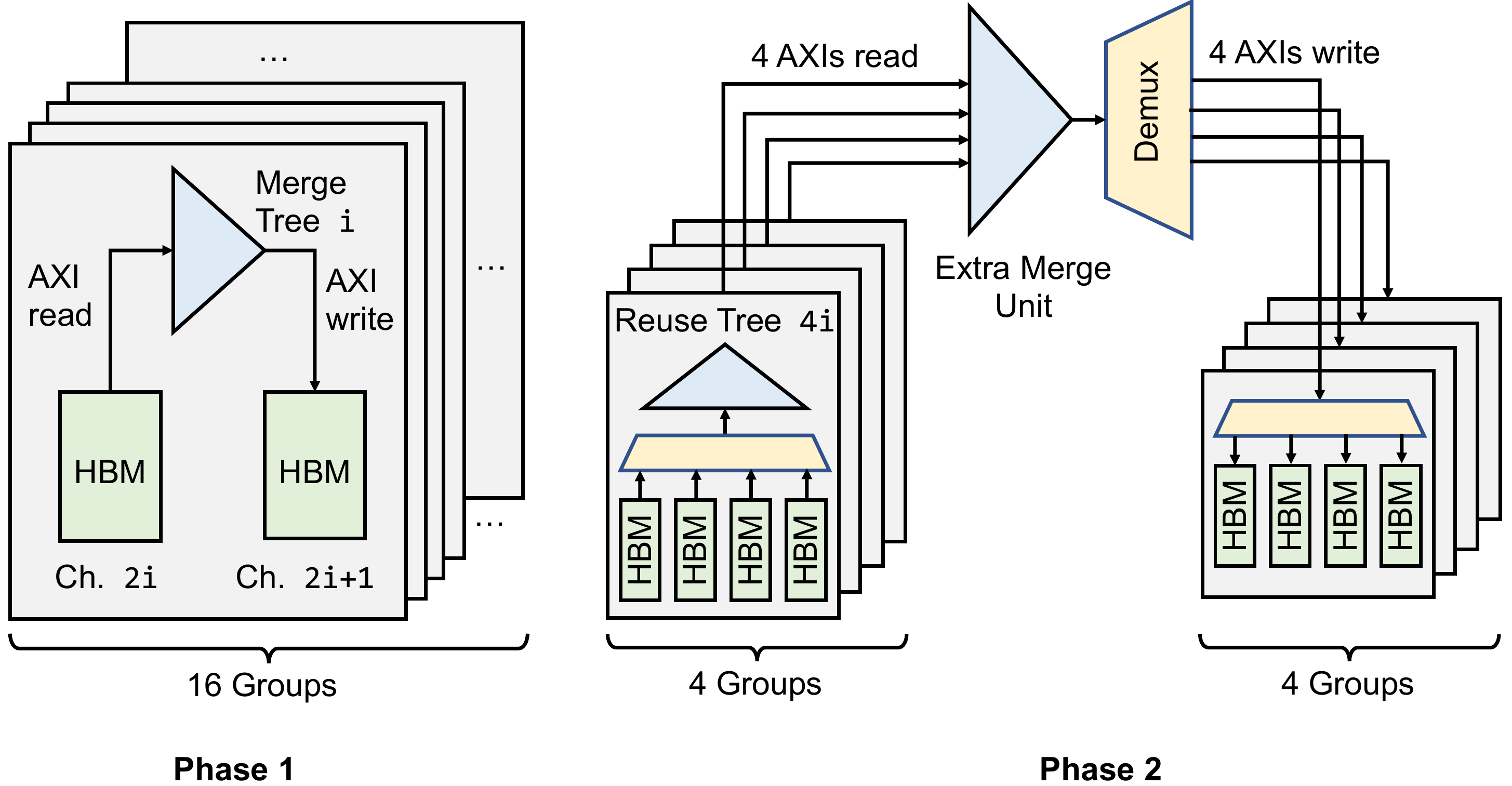}
    \caption{The overall architecture of the two phases in \toolName. The merge trees rely on AXI interfaces to access HBM channels. The role of demux is covered in Section~\ref{subsec:memory_write}. 
    }
    \label{fig:microarch}
\end{figure}

The overall architecture of \toolName~is shown in Figure~\ref{fig:microarch}. There are 32 HBM channels available, so \toolName~uses 16 merge tree kernels in phase 1. Given a total of $N$ unsorted elements, we split them into 16 sequences, each of which contains $N/16$ elements and is stored in one HBM channel. Then each of the 16 trees will stream the unsorted sequence from one HBM channel and stream the merged (sorted) sequence to another HBM channel, as described in Section~\ref{subsec:existing_tree}. We design each tree so that it has 16 leaves and outputs 64 bytes per cycle at its root to saturate the bandwidth of a single HBM channel. The leaf number for each merge tree is chosen to be 16 is because the leaf buffers cannot exceed the available on-chip BRAM limit. Assuming each element is 64-bit, the merge tree will output 8 elements every cycle.

In phase 2, \toolName~reuses 4 merge trees from phase 1 and adds two levels of merge units to form a wider merge tree. This merge tree has 64 leaves and outputs 256-byte elements per cycle. Since there are 16 sequences sitting in 16 HBM channels after phase 1, we split each sequence into 4 segmented sub sequences. Then each of the 64 sub sequences is fed into one of the 64 leaves for one pass to get the final sorted results.

In phase 2, the input buffers and each merge unit of the four reused merge trees stay exactly the same as phase 1. We only need to change the memory addresses and the length of the sequences that the AXI interfaces read from each channel for each of the 64 leaves. The change of the write behavior of phase 2 is discussed in Section~\ref{subsec:memory_write}. 

\subsection{Tuning Sorted Sequence Size from Phase 1}
\begin{figure}[!t]
    \centering
    \includegraphics[width=\columnwidth]{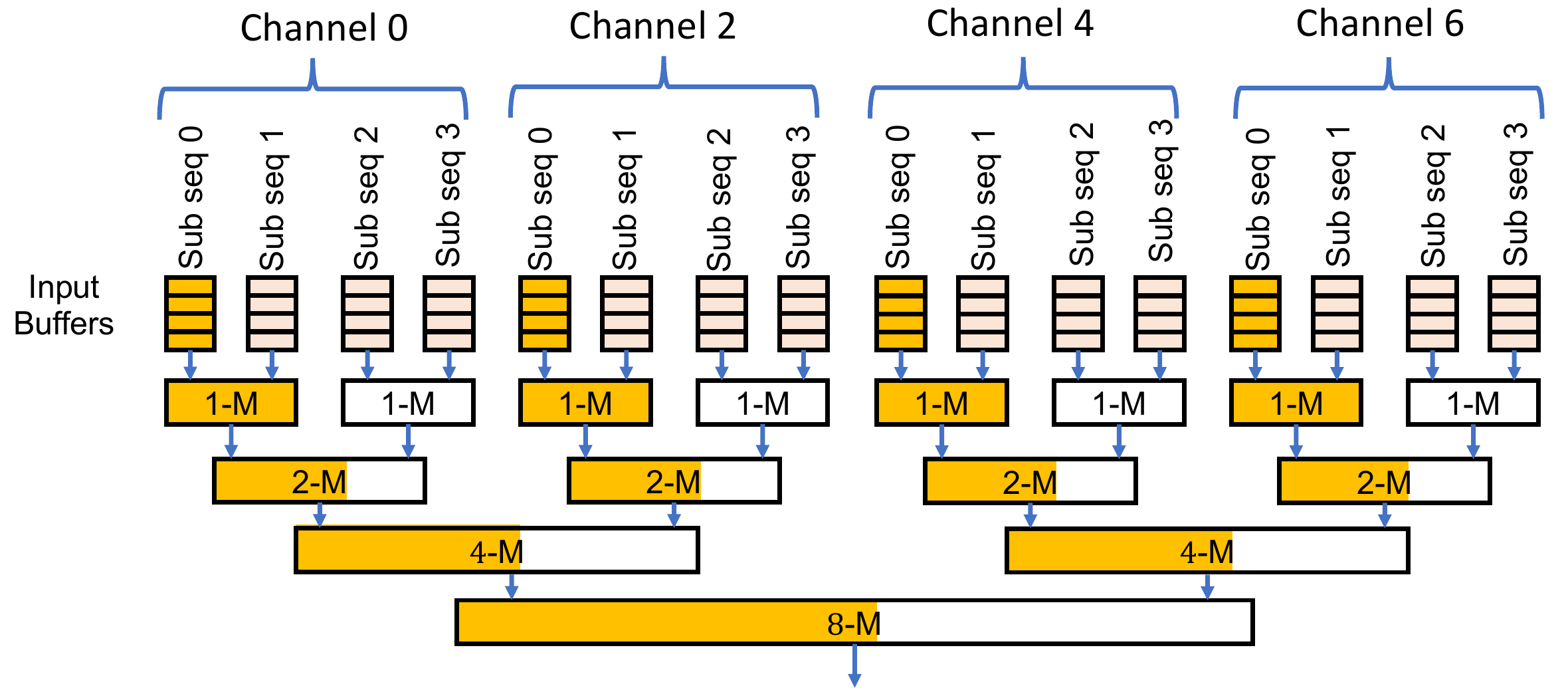}
    \caption{The active rate of one reused tree in phase 2, if the sequence in each channel is completely sorted. Initially, only the leaves feeding sub sequence 0 from each channel are active, then are the leaves feeding sub sequences 1, and so on. Leaves marked as yellow are actively feeding. Merge units partially marked mean they are idle half of the time.}
    \label{fig:active_merge_tree}
\end{figure}

One hidden requirement for a merge tree to output $p$ elements per cycle is that, on average, it has $p$ leaves out of its total $l$ leaves providing elements every cycle. Section~\ref{subsec:arch} mentions that each of the 16 sequences in 16 HBM channels after phase 1 needs to be divided into 4 segmented sub sequences so that there are 64 sub sequences fed into the 64 leaves of the tree in phase 2. If each sequence is totally sorted after phase 1, then the 4 sub sequences have the relation that elements in sub sequence 0 are always smaller than elements in sub sequence 1, and so on. When such 4 sub sequences are fed into 4 leaves, only 1 of the 4 leaves will feed the element to the merge tree at any cycle. For each of the reused merge trees that has 16 leaves, this means only 4 leaves are providing elements into the tree per cycle, as shown in Figure~\ref{fig:active_merge_tree}. As a result, each merge tree is idle half of the time. Although each tree can output 8 elements per cycle, the average throughput of the tree is 4 elements per cycle, which is below our expectation.


To solve this issue, we need to tune the sorted sequence size from phase 1. Instead of completely sorting the $N/16$-element sequence in each channel, we want 4 $N/64$-element sorted sub sequences from phase 1. This is done by controlling the number of elements that go into the leaves in the last pass of phase 1. Since the sorted sequence size always grows by the number of leaves $l$, we deliver $N/1024$ elements into each leaf. At the tree root side, we get $N/64$-element sorted sequence. We repeat this process 4 times for each merge tree in phase 1 to get 4 $N/64$-element sorted sub sequences.

\begin{figure}[!t]
    \centering
    \includegraphics[width=0.8\columnwidth]{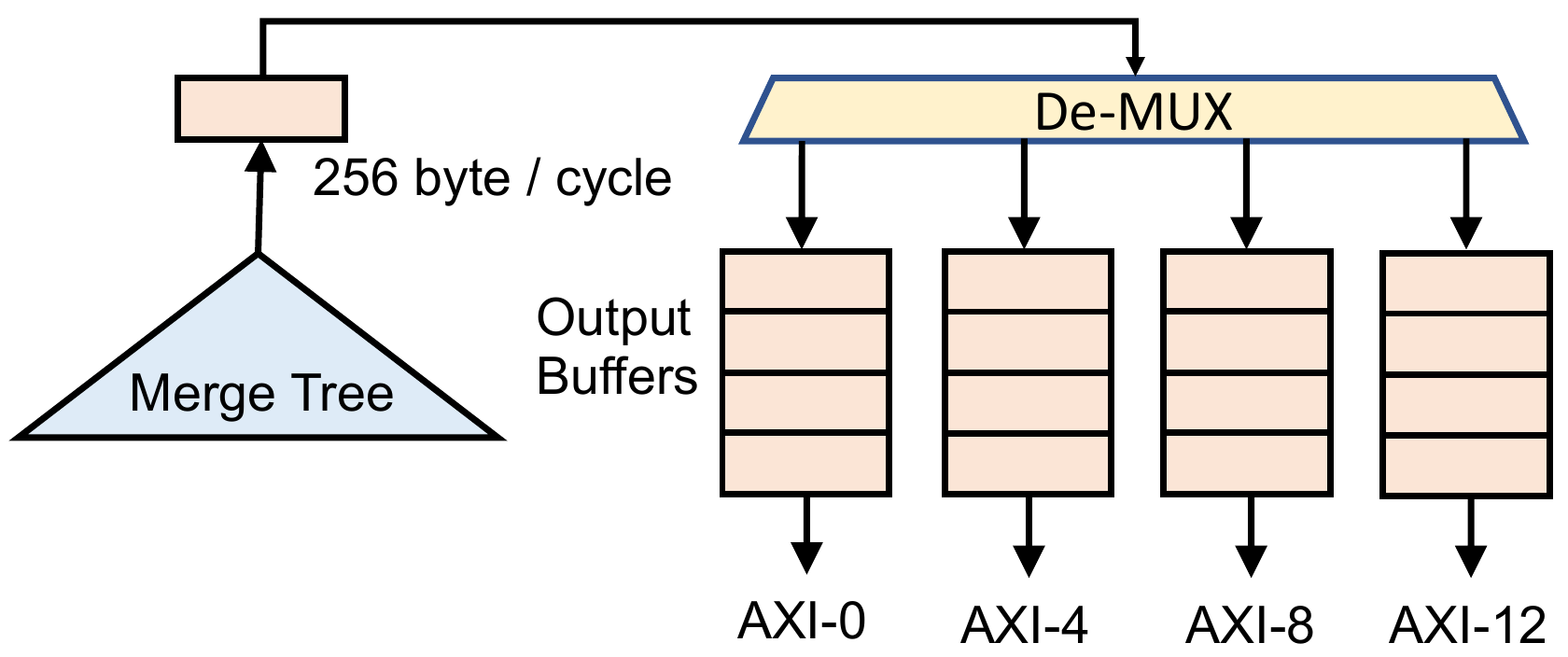}
    \caption{\toolName's write behavior of phase 2. Writes from 4 AXI interfaces are used to match the tree's throughput.}
    \label{fig:write_phase2}
\end{figure}

\subsection{Memory Write Pattern in Phase 2} \label{subsec:memory_write}
Since the 4$\times$ wider merge tree of phase 2 outputs 256-byte sorted elements per cycle, there are 4 AXI interfaces writing to the HBM channels in parallel. The choice of the specific AXI interfaces will be explained in Section~\ref{subsec:data_layout}.  The on-chip memory is not big enough to hold all the continuously sorted elements before they are written back to entirely fill one HBM channel. Therefore, these continuously sorted elements have to be split into batches and then written back to separate HBM channels through 4 AXI interfaces. Figure~\ref{fig:write_phase2} depicts the details of such behavior. In our implementation, we choose the batch size to be 4 KB, e.g., we write the first 4 KB sorted elements into the first output buffer, then write the second 4 KB sorted elements into the second output buffer, and so on. The elements in the four buffers will then be written to HBM channels via the four AXI interfaces. As a result, although the elements are completely sorted, one needs to pick the 4 KB batches channel by channel to form the final sorted results.

%% file: 4-hbm_optimize.tex
\section{HBM-Specific Optimizations} \label{sec:hbm_profile}

In this section, we introduce the design choices in \toolName~that are coupled with specific HBM characteristics. These HBM-related optimizations are necessary to achieve a full bandwidth usage. First, we minimize the number of AXI interfaces to reduce the resource overhead. Second, we propose a dedicated data layout with spatial locality to prevent bandwidth degradation. In addition, we properly set different burst sizes for different merge trees to strike a balance between the area overhead and the bandwidth. Finally, we floorplan and pipeline our design based on the architecture of the HBM platform for timing optimization. 

\subsection{Avoiding Unnecessary AXI Interfaces} \label{subsec:axi_conv}

According to our measurement, an AXI rate converter in Figure~\ref{fig:hbm_axi} for an HBM channel requires about 5K LUTs and about 6K Flip-Flops (FFs) on the Xilinx U280 FPGA. As a result, the naive choice of instantiating one AXI interface for each of the 32 HBM channels will result in about 320K LUTs, which is more than 40\% of the available LUTs on the bottom die where the HBM resides. Such high resource overhead will squeeze the space for user logic and cause severe routing issues.

To address this issue, we utilize the internal crossbar among HBM channels. For each merge tree of phase 1, we only need one AXI module to read from and write to two adjacent HBM channels.
This way halves the AXI area overhead and effectively reduces the routing congestion in the bottom die of the FPGA.





\subsection{Data Layout Optimization} \label{subsec:data_layout}

Although each AXI interface from the user side can access any of the 32 HBM channels, non-local data accesses could cause contention over the lateral connections between channel groups, thus leading to bandwidth decrease. To access an HBM channel outside the group, the data must occupy and traverse through each of the lateral connections until it reaches the destination. Two memory accesses requiring the same lateral connection will cause a conflict and one will be blocked.  

Therefore, we must carefully arrange for the data placement to minimize out-of-group channel accesses and avoid conflicts in lateral connections. Table~\ref{tab:axi_bundle} summarizes the target HBM channels for each AXI interface. In the first phase, merge tree $i$ uses AXI-$i$ to only read/write a pair of nearby HBM channels $2i$ and $2i+1$, where $i$ ranges from 0 to 15. In the second phase, the merge tree $4i$ uses AXI-$4i$ to read/write between channels $8i+2j$ and channels $8i+2j+1$, where $i$ \& $j$ both range from 0 to 3. 

\input{Table/axi_bundle}

\begin{figure}[!t]
    \centering
    \includegraphics[width=0.8\columnwidth]{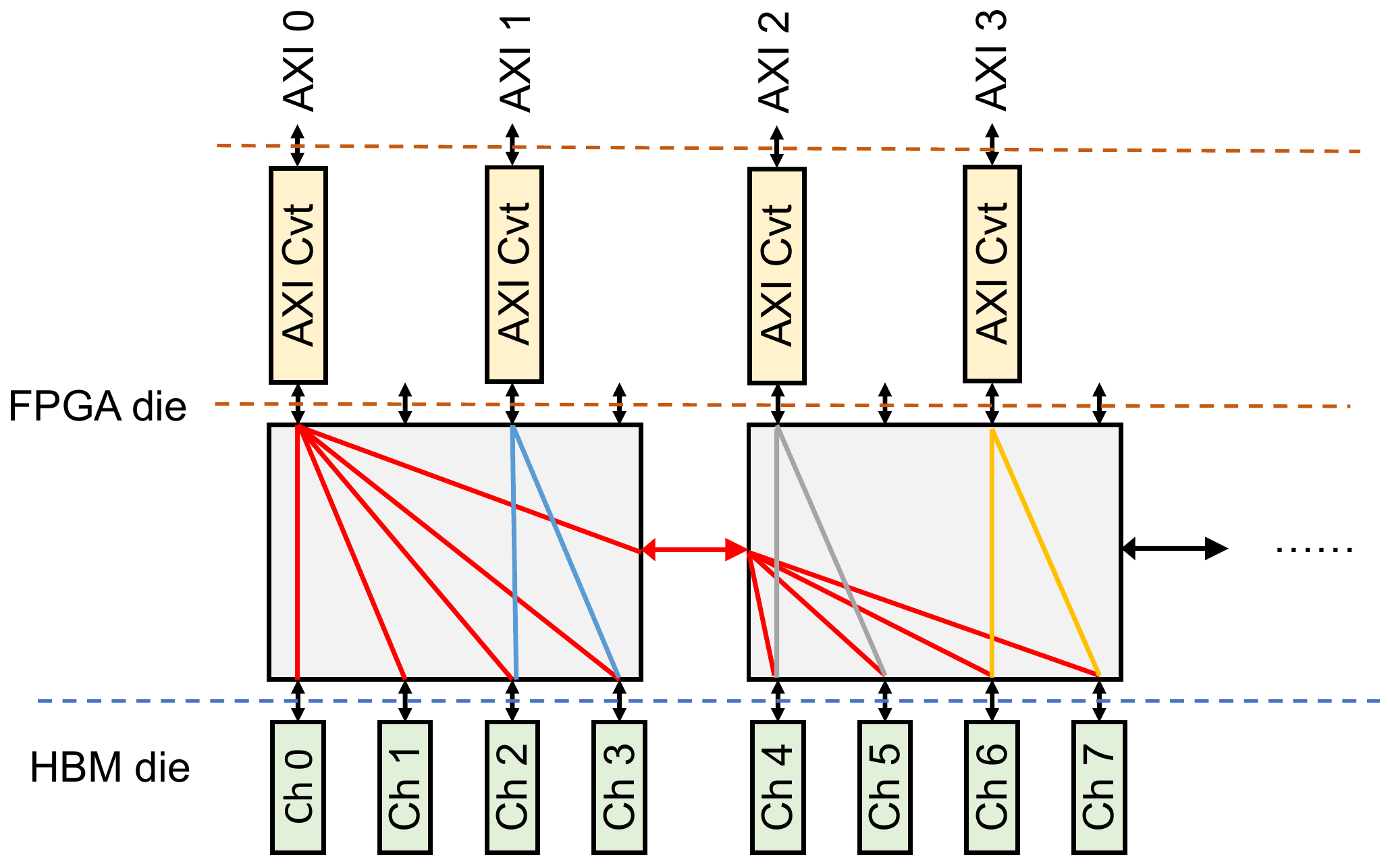}
    \caption{Usage of the first 4 AXI interfaces. The same behavior is repeated for rest of the 12 AXI interfaces.}
    \label{fig:axi_detail}
\end{figure}

Figure~\ref{fig:axi_detail} illustrates the access pattern of the first four AXI interfaces in different colors. AXI-$1$, $2$ and $3$ will only access the two channels within its crossbar group. Although the AXI-$4i$ in the second phase needs to access channels outside of its local group, our data layout guarantees that different AXI interfaces will not cause conflicts in lateral connections.

\subsection{Burst Size Optimization}\label{subsec:hbm_chan}

For \toolName, each of the AXI interfaces will always access the HBM in burst mode. We need to properly choose the burst sizes because a large burst requires excessive buffers while a small burst may not fully utilize the bandwidth.

In order to select proper burst sizes, we profile the HBM performance based on the memory access pattern of each AXI interface. As shown in Fig.~\ref{fig:axi_detail}, In phase one, each AXI reads from one channel and writes to the adjacent channel; In phase two, AXI-$0$ reads from four even/odd channels in a round-robin way and writes to the other four nearby odd/even channels.

\begin{figure}[!t]
    \centering
    \includegraphics[width=\columnwidth]{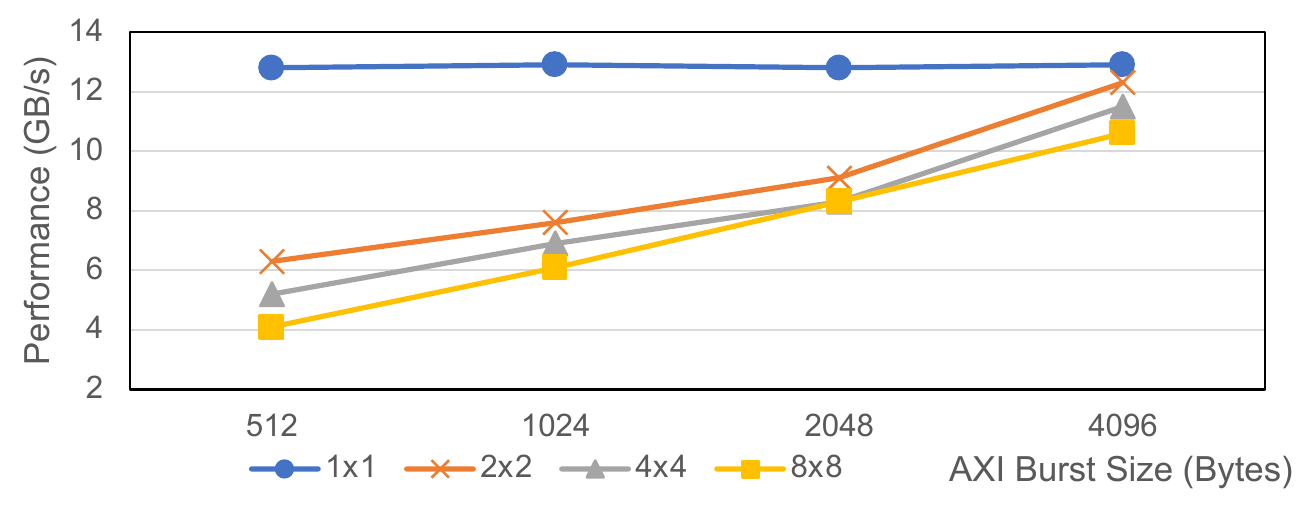}
    \caption{HBM channel performance of inter-crossbar \& intra-crossbar behaviors when varying the AXI burst size. The number of outstanding bursts is fixed at 32.}
    \label{fig:chan_perf}
\end{figure}

We define reading from $m$ HBM channels and writing to their nearby $m$ channels as pattern $m$$\times$$m$. The patterns of phase 1 and phase 2 are thus 1$\times$1 and 4$\times$4 by this definition. For completeness, we also perform tests for patterns 2$\times$2 and 8$\times$8. Then we measured the achieved bandwidth when using different AXI burst sizes as seen in Figure~\ref{fig:chan_perf}. 

Based on our experiments, any AXI burst size from 512 B to 4 KB can achieve the peak HBM bandwidth for the 1$\times$1 pattern, which is used by merge trees in phase 1. However, for the 4$\times$4 inter-crossbar memory access pattern used in phase 2, the AXI burst size needs to be maximized to  4 KB to achieve the best bandwidth. Therefore, we set the burst size to be 1 KB for the 12 merge trees that are not reused in phase 1, and 4 KB for the 4 merge trees that are reused in phase 2. 

Minimizing the burst sizes under the bandwidth constraints could significantly save resources because not all buffers are implemented in BRAM. For each merge tree, each leaf node requires a 512-bit wide input buffer (see Figure~\ref{fig:single_merge_tree}) to hold the read bursts from AXI interfaces. When implemented using BRAM, each buffer consumes at least 7.5 BRAMs on Xilinx FPGAs. Since we have 16 merge trees and each tree has 16 leaves, the FPGA does not have enough BRAMs to implement all the buffers. So, about $3/4$ of the buffers are implemented using LUT-based shift registers. While BRAM-based buffers are insensitive to the burst size, the area of a LUT-based buffer is directly proportional to the buffer depth. A single LUT can implement a 1-bit shift register with a depth of 32 and 512 LUTs can hold two AXI bursts of 1 KB. Reducing the bursts size from 4 KB to 1 KB directly reduces the LUTs consumption by 4$\times$. Further reducing the burst size from 1 KB to 512 B does not save more LUTs because 2 bursts of 512 B still require 512 LUTs. The actual sorting results with different burst sizes in \toolName~is further discussed in Section~\ref{subsec:exp_burst}.

%% file: Table/axi_bundle.tex
\begin{table}[!htb]
    \centering
    \caption{Correspondence between each user side AXI and the HBM channels it accesses in \toolName, $i$=0-3.}
    \label{tab:axi_bundle}
    \resizebox{0.6\columnwidth}{!}{
    \begin{tabular}{cc}
    \hline
    \multicolumn{1}{|c|}{AXI NO.} & \multicolumn{1}{c|}{HBM Channel NO.} \\ \hline
    \multicolumn{1}{|c|}{AXI-$4i$} & \multicolumn{1}{c|}{8$i$ - (8$i$+7)} \\ \hline
    \multicolumn{1}{|c|}{AXI-$(4i+1)$} & \multicolumn{1}{c|}{(8$i$+2) - (8$i$+3)} \\ \hline
    \multicolumn{1}{|c|}{AXI-$(4i+2)$} & \multicolumn{1}{c|}{(8$i$+4) - (8$i$+5)} \\ \hline
    \multicolumn{1}{|c|}{AXI-$(4i+3)$} & \multicolumn{1}{c|}{(8$i$+6) - (8$i$+7)} \\ \hline
    \end{tabular}
    }
\end{table}

%% file: 5-floorplan.tex
\subsection{Floorplanning Optimization} \label{subsec:floorplan}

\begin{figure}[!t]
    \centering
    \includegraphics[width=0.7\columnwidth]{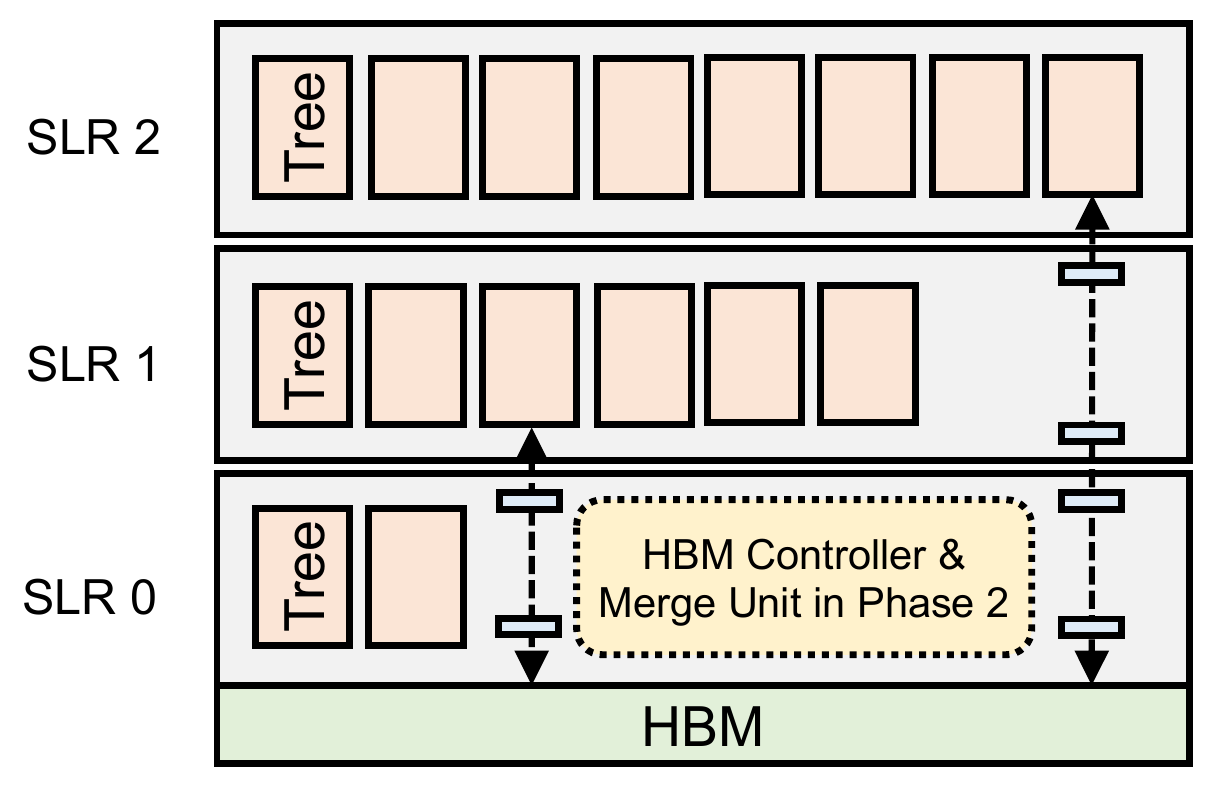}
    \caption{The floorplan of \toolName~on the Xilinx U280 FPGA. Each Super Logic Region (SLR) is a single FPGA die slice in the Stacked Silicon Interconnect (SSI) FPGA device}
    \label{fig:floor_plan}
\end{figure}

Thanks to the parallel sorting in the first phase and merge tree reuse in the second phase, we are able to separate each of the merge tree kernels independently and apply a simple yet efficient resource model to better distribute these kernels across the three FPGA dies, as shown in Figure~\ref{fig:floor_plan}. Please note that the merge tree kernels in the first phase only expose simple AXI interfaces and each of the merge tree has the same size. This feature allows us to easily implement a coarse-grained floorplanning with the granularity of the separate merge tree kernel. In contrast, if we simply scale the single merge tree and want to make use of all of the FPGA dies, it would requires significantly more engineering efforts to manually split the design logics and fit them into each die. 

Considering that the sorted output from the merge tree of phase 2 will be directly written to the HBM channels, a natural choice is to place the extra merge units on the bottom die to minimize the signal crossing. After that, we will migrate as many merge tree kernels to the mid and top dies, to save more spaces in the bottom die to alleviate the routing issues. 

We derive a simple resource model to solve this floorplanning problem. Let $L$ be the resource consumption of each merge tree, $u_{i}$ be the number of merge trees and $a_{i}$ be the available resources on the $i$-th die, $w$ be the width of an AXI interface, and $W$ be the available signals crossing two neighbor dies. Then the problem can be formulated as \emph{maximizing} $(u_{1}+u_{2})$ under the following constrains:
\begin{align*}
\begin{cases}
    (u_{1}+u_{2}) \cdot w \leq W \quad \ \ \  (1) \\
    u_{1} \cdot L \leq a_{1} \quad \quad  \quad \quad \quad \  (2) \\ 
    u_{2} \cdot L \leq a_{2} \quad \quad  \quad \quad \quad \  (3) 
\end{cases}
\end{align*}
where constraint (1) limits the AXI connections from all merge trees in both die 1 and 2 to die 0 do not exceed the available amount of signals crossing die 0 and 1, and constrains (2) and (3) make sure the merge trees placed on die 1 and 2 do not exceed each die's available resource. 

By solving the above problem, we place eight merge trees on the top die and six merge trees on the middle die. There are another two merge trees remaining on the bottom die; we select these two merge trees as merge tree 0 and merge tree 8. The detailed layout are presented in Section~\ref{subsec:exp_layout}.

Finally, to improve the timing, the signals crossing dies need to be pipelined with registers~\cite{fpga'21-autobridge}. In general, we add 2 stages of pipeline registers when a signal crosses one die and add 4 stages of pipeline registers when it crosses two dies.

%% file: 6-experiment.tex
\section{Experimental Results} \label{sec:experiment}

\subsection{Experimental Setup} \label{subsec:exp_setup}
We perform all of the experiments on the Xilinx U280 FPGA board. The kernel is developed using System Verilog and is synthesized and implemented using Xilinx Vitis 2020.2. We use the pblock method~\cite{pblock} to place the design modules during the floorplanning. The input data are in a key-value pair format, each with a 32-bit key and a 32-bit value. The keys used in the experiments have a uniform distribution. To ensure the keys are fully randomized, we first generate $N$ records whose keys are incremented from 1 to $N$, then we use the $random\_shuffle()$ function from the python library to shuffle the records. The sorted results are validated by checking whether their keys are from 1 to $N$. 

\subsection{\toolName~Configuration \& Resource Utilization} \label{subsec:exp_config}
We implement 16 merge trees in total, each with 16 leaves and outputs 8 elements (each 8 bytes) per cycle.
In phase 1, \toolName~uses all 16 merge trees. In phase 2, \toolName~reuses 4 of the merge trees to form a wider merge tree that can sort 32 elements (each 8 bytes) per cycle. 

The resources utilization of the dynamic region\footnote{The FPGA is partitioned into two regions, with the dynamic region reserved for user logic and the static region used for infrastructure IPs.} is listed in Table~\ref{tab:resource_overall}, including our \toolName~kernel and HBM controllers. \toolName~consumes 54.6\% LUTs, 35.5\% FFs and 56\% BRAMs. 

Table~\ref{tab:resource_phase} shows the resource breakdown of each type of the merge trees. For those trees that are not reused (e.g., merge tree 1), each of them takes less than 3\% of the available resources. This is consistent with the observation in Section~\ref{subsec:analysis_single_tree} that it takes significantly less resources for a single merge tree to saturate one HBM channel than multiple HBM channels. Meanwhile, a reused tree (e.g., merge tree 4) requires more LUTs because its input buffers need to accommodate larger data bursts. As for the extra 3 merge units used in phase 2, they takes much more resources than a single merge tree, which validates the theoretical analysis that directly scaling the tree throughput requires the resources to grow superlinearly.

\input{Table/overall_resources}

\input{Table/phases_resources}

\subsection{Design Layout \& Frequency} \label{subsec:exp_layout}

\begin{figure}[!t]
    \centering
    \includegraphics[width=0.7\columnwidth]{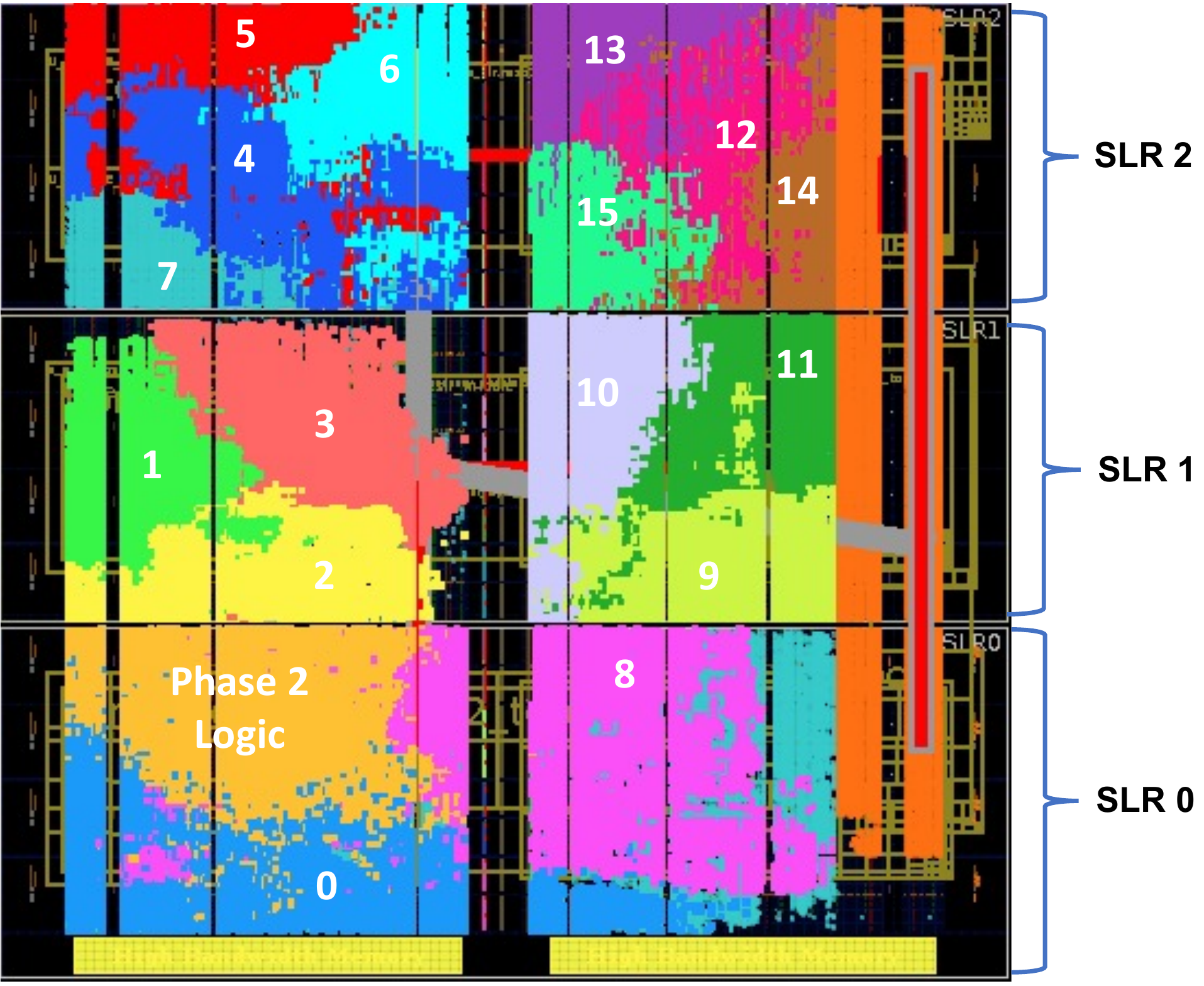}
    \caption{Layout of \toolName~implemented on the Xilinx U280 FPGA board. Number 0-15 label the 16 merge trees of phase 1. The orange part in the bottom die represents the extra logic used to form the wider merge tree of phase 2.}
    \label{fig:die}
\end{figure}

Figure~\ref{fig:die} shows the final placements of all merge trees and the extra phase 2 logic. Our floorplanning and pipelining effectively reduces local routing congestion, so that \toolName~could achieve 214 MHz in user clock and 414 MHz in the HBM control clock. Without our floorplanning, Vivado failed in routing even with the highest optimization level.



\begin{figure}[!t]
    \centering
    \includegraphics[width=\columnwidth]{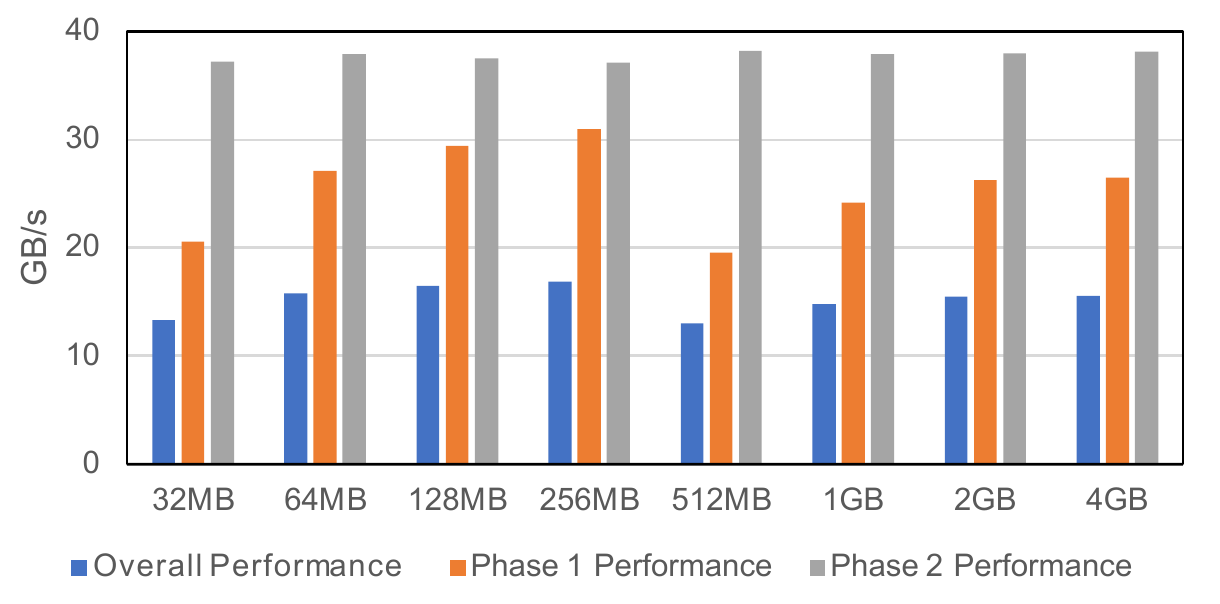}
    \caption{Sorting performance with different data sizes. The overall performance is calculated through dividing the data size  by the total sorting execution time.}
    \label{fig:sort_perf}
\end{figure}

\input{Table/perf_comp}

\subsection{Overall Sorting Performance} \label{subsec:exp_perf}
Figure~\ref{fig:sort_perf} presents the performance of sorting data ranging from 32 MB to 4 GB in size. The maximal data size that we can handle is half of the HBM capacity, which is 4 GB in the Xilinx U280 FPGA. This is because half of the HBM channels are used for reading and the other half for writing. When sorting 4 GB of data, \toolName~achieves an overall performance of 15.6 GB/s. Please note that 4 GB is half of the HBM capacity and the maximal size that we are able to sort. Sorting 4GB data has a performance of 26.5 GB/s in the first phase and 38 GB/s in the second phase. 

Figure~\ref{fig:sort_perf} also shows a performance drop when the date size is increased from 256 MB to 512 MB. This is because sorting 512MB data takes 6 passes in phase 1 while sorting 256 MB data takes 5 passes. Besides, sorting 512 MB to 4 GB data requires the same number of passes in phase 1, but the sorting performance still varies, which is similar for sorting 32 MB to 256 MB data. This is because some input leaves of the merge trees may not be fully active in the last pass of phase 1, similar to what is illustrated in Figure~\ref{fig:active_merge_tree}.

Based on our measurement, we are able to utilize at least 318 GB/s of HBM bandwidth in the first phase, where we run 6 passes and each pass reads and writes to HBM channels simultaneously. Thus the average HBM bandwidth utilized is at least $26.5 \times 6 \times 2 =318$ GB/s. In fact, the used bandwidth is even higher because we have not account for the idle time of the merge units, which needs to be reset after merging two sorted sequences. In the initial pass, the length of the sorted sequences is 1, so the merge units have to be reset frequently.

\subsection{Sorting Performance with Different Burst Sizes} \label{subsec:exp_burst}

\begin{figure}[!t]
    \centering
    \includegraphics[width=\columnwidth]{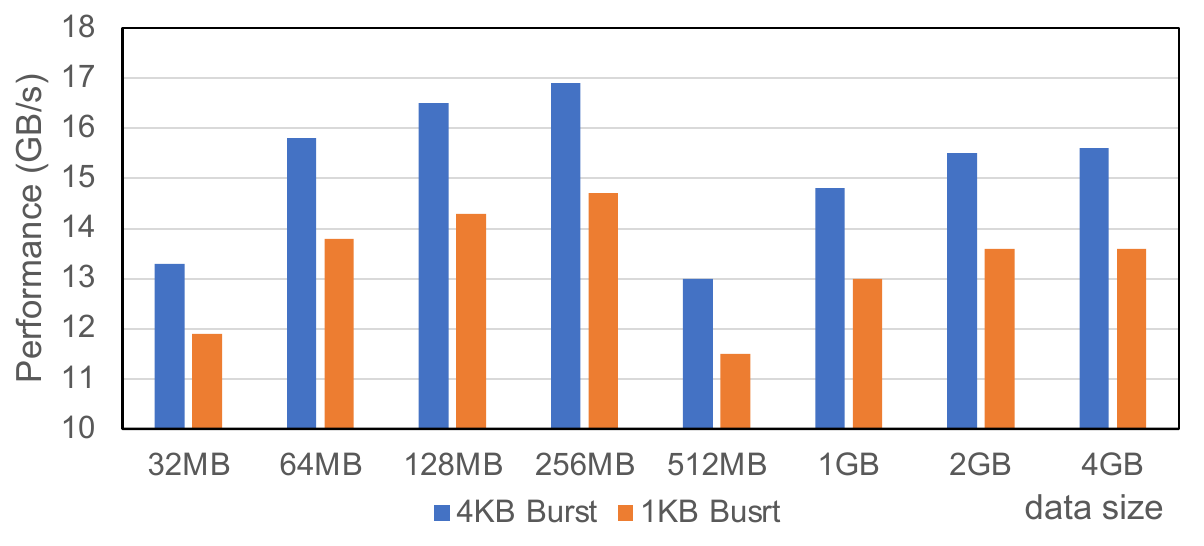}
    \caption{Overall sorting performance with the same AXI burst size in phase one but different AXI burst sizes in phase two.}
    \label{fig:sort_perf_burst}
\end{figure}

Figure~\ref{fig:sort_perf_burst} compares the performance of two burst size choices. Both tests set the burst size as 1 KB in phase 1. In the second phase, we set the burst sizes to be 1 KB and 4 KB. 
The results show that reading in 4KB bursts in phase 2 gives better performance, which matches the profiling results in Section~\ref{subsec:hbm_chan} that the designers need to use 4KB AXI bursts to maximize the HBM channels' bandwidth when they use one AXI to access multiple HBM channels.

\subsection{Comparison with State-of-the-art CPU \& FPGA Sorters} \label{subsec:exp_cmp_fpga}
\toolName~is an in-memory sorter~\cite{knuth}, which stores the entire input in main memory such as an HBM or a DRAM. An in-memory sorter has much higher overall sorting performance than an in-storage sorter, which requires a second-level storage such as SSD to sort a larger dataset and thus is bounded by the slow I/O~\cite{fccm'17-tera,fccm'21}. In fairness, we compare \toolName~against several CPU and FPGA in-memory sorters here.
The state-of-the-art CPU in-memory sorter in~\cite{vldb15-radix} can sort 4GB of data in 1.7 second, achieving an overall sorting performance of about 2.3 GB/s. Compared to it, \toolName~achieves 6.7$\times$ speedup.

Table~\ref{tab:perf_comp} lists the existing FPGA in-memory sorters. The best-performing DRAM-based FPGA sorter in~\cite{isca'20} can fully utilize 4 DDR4 DRAM channels on the AWS F1 datacenter FPGA. When sorting 4 GB data, it achieves an overall sorting performance of 7.1 GB/s. Comparing to it, \toolName~achieves 2.2$\times$ speedup. Besides, we also list the absolute merge tree throughput, which is the number of elements ($p$) that the merge tree can output per cycle multiplied by the design frequency and the element's width in bytes, to show the utilization of the off-chip memory bandwidth of different merge tree designs. \toolName~has an entire tree throughput of 219 GB/s in phase 1 and saturates the 420 GB/s HBM bandwidth (HBM is half-duplex so half of its bandwidth is for reading data and the other half is for writing data). In phase 2, \toolName~still has 1.7$\times$ and 14$\times$ higher merge tree throughput than that of ~\cite{isca'20,fpl'20}, respectively.

One may find that the overall sorting performance improvement of \toolName~over~\cite{isca'20} is much less than the 6$\times$ off-chip bandwidth increase. This is because the on-chip resources of the Xilinx U280 FPGA is merely enriched by about 20\% compared to the AWS F1 FPGA. As analyzed in Section~\ref{sec:background}, the on-chip resources become the new bottleneck to linearly scale the overall sorting performance. While \toolName~could address this issue to some extent, its overall performance is still bound by the merge tree in phase 2. That is why only 2.2$\times$ speedup is reported.


\footnotetext[3]{~\cite{fpl'20} reported 49$\times$ speedup over the C++ std::sort() implementation on an ARM Cortex A53 core. We re-implement C++ std::sort() on a more advanced Intel Xeon E5-2680 core and multiply the performance by 49$\times$ to get the listed value.}

\subsection{Comparison with Scaled Single Merge Tree} \label{subsec:exp_cmp_single}
We also compare \toolName~to a single giant merge tree, which is scaled to output 32 elements per cycle and has a throughput equal to writing 4 HBM channels, i.e., the same as the merge tree throughput of phase two in \toolName. Unfortunately, we are not able to route this single merge tree, as it requires significantly more engineering efforts of manual placement optimization compared to \toolName. The reason why it is easier to floorplan \toolName~than to floorplan single merge tree is explained in Section~\ref{subsec:floorplan}. Nonetheless, we can still estimate its performance, assuming it achieved the same design frequency as \toolName. 

The best estimation is that the single merge tree has the same number of leaves as the sum of the 16 merge tree leaves in phase one of \toolName, which is 256. This means sorting 4 GB data still requires 4 passes. Since the throughput of the merge tree is the same as \toolName's throughput in phase 2, which is 38 GB/s, the overall sorting performance of the single merge tree is thus at most 38/4 = 9.5 GB/s, much less than the 15.6 GB/s achieved in \toolName. 


%% file: Table/overall_resources.tex
\begin{table}[!htb]
    \centering
    \caption{Resources utilization and percentage of the \toolName~kernel and HBM controllers. DSPs are used to calculate the addresses of AXI read and write.}
    \label{tab:resource_overall}
    \resizebox{\columnwidth}{!}{
    \begin{tabular}{ccccc}
    \hline
    \multicolumn{1}{|c|}{Components} & \multicolumn{1}{c|}{LUTs} & \multicolumn{1}{c|}{Flip-Flops} & \multicolumn{1}{c|}{BRAMs} & \multicolumn{1}{c|}{DSPs} \\ \hline
    \multicolumn{1}{|c|}{Available} & \multicolumn{1}{c|}{1,066,848} & \multicolumn{1}{c|}{2,144,089} & \multicolumn{1}{c|}{1,487} & \multicolumn{1}{c|}{8,484} \\ \hline\hline
    \multicolumn{1}{|c|}{\toolName~kernel} & \multicolumn{1}{c|}{582,244} & \multicolumn{1}{c|}{750,505} & \multicolumn{1}{c|}{834} & \multicolumn{1}{c|}{84} \\ \hline
    \multicolumn{1}{|c|}{Percentage} & \multicolumn{1}{c|}{54.6\%} & \multicolumn{1}{c|}{35.5\%} & \multicolumn{1}{c|}{56.0\%} & \multicolumn{1}{c|}{1.0\%} \\ \hline\hline
    \multicolumn{1}{|c|}{HBM Controllers} & \multicolumn{1}{c|}{87,848} & \multicolumn{1}{c|}{113,814} & \multicolumn{1}{c|}{4} & \multicolumn{1}{c|}{0} \\ \hline
    \multicolumn{1}{|c|}{Percentage} & \multicolumn{1}{c|}{8.2\%} & \multicolumn{1}{c|}{5.3\%} & \multicolumn{1}{c|}{0.2\%} & \multicolumn{1}{c|}{0.0\%} \\ \hline
    \end{tabular}
    }
\end{table}

%% file: Table/phases_resources.tex
\begin{table}[!htb]
    \centering
    \caption{Resources utilization of individual merge trees. Here we pick one merge tree from each type since they have slightly different resource consumption.}
    \label{tab:resource_phase}
    \resizebox{\columnwidth}{!}{
    \begin{tabular}{ccccc}
    \hline
    \multicolumn{1}{|c|}{Components} & \multicolumn{1}{c|}{LUTs} & \multicolumn{1}{c|}{Flip-Flops} & \multicolumn{1}{c|}{BRAMs} & \multicolumn{1}{c|}{DSPs} \\ \hline\hline
    \multicolumn{1}{|c|}{Merge tree 1} & \multicolumn{1}{c|}{28,788} & \multicolumn{1}{c|}{39,905} & \multicolumn{1}{c|}{37.5} & \multicolumn{1}{c|}{2} \\ \hline
    \multicolumn{1}{|c|}{Percentage} & \multicolumn{1}{c|}{2.7\%} & \multicolumn{1}{c|}{1.9\%} & \multicolumn{1}{c|}{2.5\%} & \multicolumn{1}{c|}{0.02\%} \\ \hline\hline
    \multicolumn{1}{|c|}{Merge tree 4} & \multicolumn{1}{c|}{39,195} & \multicolumn{1}{c|}{39,459} & \multicolumn{1}{c|}{60} & \multicolumn{1}{c|}{10} \\ \hline
    \multicolumn{1}{|c|}{Percentage} & \multicolumn{1}{c|}{3.7\%} & \multicolumn{1}{c|}{1.8\%} & \multicolumn{1}{c|}{4.0\%} & \multicolumn{1}{c|}{0.1\%} \\ \hline\hline
    \multicolumn{1}{|c|}{Extra logic of phase 2} & \multicolumn{1}{c|}{60,239} & \multicolumn{1}{c|}{102,864} & \multicolumn{1}{c|}{144} & \multicolumn{1}{c|}{20} \\ \hline
    \multicolumn{1}{|c|}{Percentage} & \multicolumn{1}{c|}{5.6\%} & \multicolumn{1}{c|}{4.8\%} & \multicolumn{1}{c|}{9.7\%} & \multicolumn{1}{c|}{0.2\%} \\ \hline
    \end{tabular}
    }
\end{table}

%% file: Table/perf_comp.tex
\begin{table*}[t]
    \centering
    \caption{Comparison with existing FPGA based in-memory sorters.}
    \label{tab:perf_comp}
    \begin{tabular}{|c|cc|cc|c|}
    \hline
       & Algorithm & Off-Chip Memory Used & Merge Tree Throughput & Design Frequency & Overall Sorting Perf. \\ \hline
     
      FPGA'20~\cite{fpga'20} & Sample sort & 4 DDR4 channels & N/A & 250 MHz & 4.3 GB/s 
      \\ \hline
      
      FPL'20~\cite{fpl'20} & Merge tree sort & 1 DDR4 channel & 4 GB/s & 214 MHz & 1.9 GB/s \footnotemark
      \\ \hline
      
      ISCA'20~\cite{isca'20} & Merge tree sort & 4 DDR4 channels & 32 GB/s & 250 MHz & 7.1 GB/s  \\ \hline
      
      TopSort & Merge tree sort & 32 HBM channels & \begin{tabular}{c}
            219 GB/s (phase 1)
           \\ 55 GB/s (phase 2)
      \end{tabular} & 214 MHz & 15.6 GB/s
      
      \\ \hline
 
    \end{tabular}

\end{table*}

%% file: 7-related_work.tex
\section{Related Work} \label{sec:related_work}
\textbf{Merge Tree Acceleration.} The existing works involved in merge tree acceleration are split into two directions. One direction focuses on optimizing the resource consumption of the merge unit itself, at the center of which is a variation of parallel sorting networks such as bitonic or even-odd sorting networks~\cite{ fccm'17-tokyo, fccm'18, fpt'18-merger, mcsoc'19}. \toolName~benefits from these works in that using efficient merge units may always reduce the on-chip resource consumption of the merge tree. In this work, we adopt the merge units from~\cite{fccm'18}, which contains two bitonic mergers, but rewrite the whole control logic to make the merge unit has the pure streaming behavior. The other direction investigates how to design efficient merge trees that interact with various memory layers such as DRAM and storage ~\cite{fccm'17-tera,fpt'18,isca'20,fpl'20,fccm'21}. These works all scale the single tree to match the off-chip bandwidth. Our work addresses a different challenge where the bottleneck of merge tree sorting is shifted to limited on-chip resources.

\textbf{HBM-Specific Optimizations.}
As for HBM-specific optimizations, \cite{fpga'22-spmv} presents an HLS design that applies a similar floorplanning strategy and achieves a design frequency of 237 MHz when using 18 HBM channels. The majority of the recent HBM-based accelerators \cite{iccad'21, arxiv'21-serpens,fpga'22-sssp} are HLS designs but are not able to get more than 190 MHz and use more than 28 HBM channels. \cite{fpt'21} implements a hash join accelerator that uses 32 HBM channels while running at 250 MHz and its random memory accesses are through 256-bit wide AXI interfaces. However, using a 256-bit wide AXI interface can only utilize half of the available HBM bandwidth at most. Compared to it, our work relies on continuous memory reads and writes, which requires the AXI interface to be 512-bit wide to maximize the memory performance and consumes significantly more on-chip resources for the AXI rate converters as well as the AXI burst buffers. Besides, we need to carefully optimize the data layout to avoid access conflicts in the lateral connections of the built-in AXI crossbars, which is never revealed in those applications with random memory access patterns.
We believe that our work provides valuable insights on how to optimize HBM-based accelerator designs, especially HLS-based designs that are struggling to fully utilize the HBM bandwidth with a high design frequency.  

%% file: 8-conclusion.tex
\section{Conclusion} \label{sec:conclusion}
In this work, we present \toolName, a high-performance two-phase sorting accelerator specialized for HBM-based FPGAs. Our analysis shows that the sorter performance on HBM-based FPGAs is bounded by the limited on-chip resources. To achieve better performance, \toolName~proposes a novel two-level sorting solution with smaller merge trees. In the first phase, \toolName~can fully utilize all the HBM bandwidth. In the second phase, \toolName~reuses the logic from the first phase to avoid the resource contention. 
Moreover, \toolName~adopts several HBM-specific optimizations to further reduce the resource overhead and improve the bandwidth utilization. Finally, it also employs coarse-grained floorplanning to achieve better time closure. \toolName~is the first HBM-based FPGA accelerator that can fully utilize all the HBM channel bandwidth and achieves 6.7$\times$ and 2.2$\times$ speedup over state-of-the-art CPU sorter and DRAM-based FPGA sorter.


%% file: 9-acknowledgement.tex
\section*{Acknowledgment}
This work is partially supported by CRISP, one of six JUMP centers and the CDSC industrial partners, including Samsung and Siemens Mentor Graphics.

%% file: biography.tex
\begin{IEEEbiography}[{\includegraphics[width=1in,height=1.25in,clip,keepaspectratio]{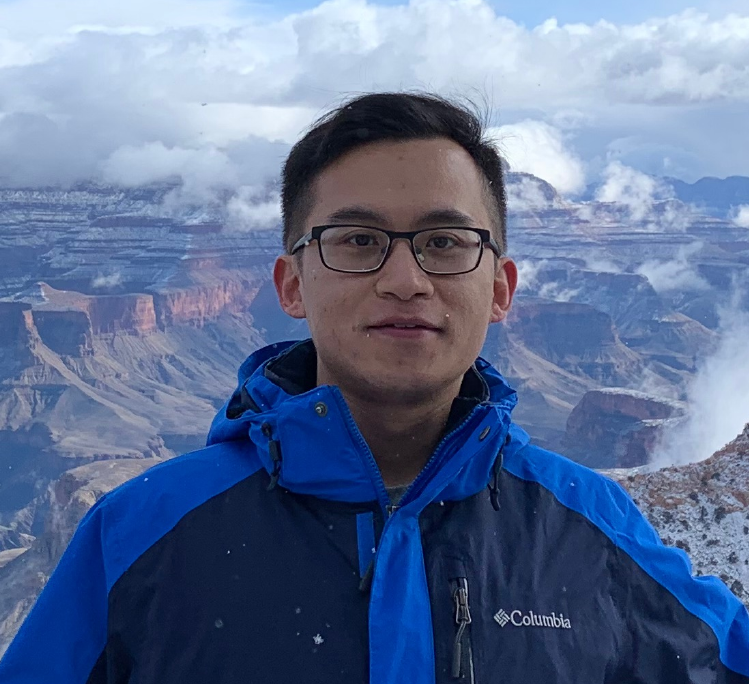}}]
{Weikang Qiao}
received his B.S. degree in Information and Communication Engineering from Zhejiang University and his M.S. degree in Electrical Engineering from UCLA. Currently, he is a fifth-year Ph.D. student in the Electrical and Computer Engineering department at UCLA. His research focuses on customized accelerator architecture designs and performance
modeling across various memory hierarchies, such as DRAM, High-bandwidth Memory (HBM) and SSDs. He is an IEEE student member.
\end{IEEEbiography}

\begin{IEEEbiography}[{\includegraphics[width=1in,height=1.25in,clip,keepaspectratio]{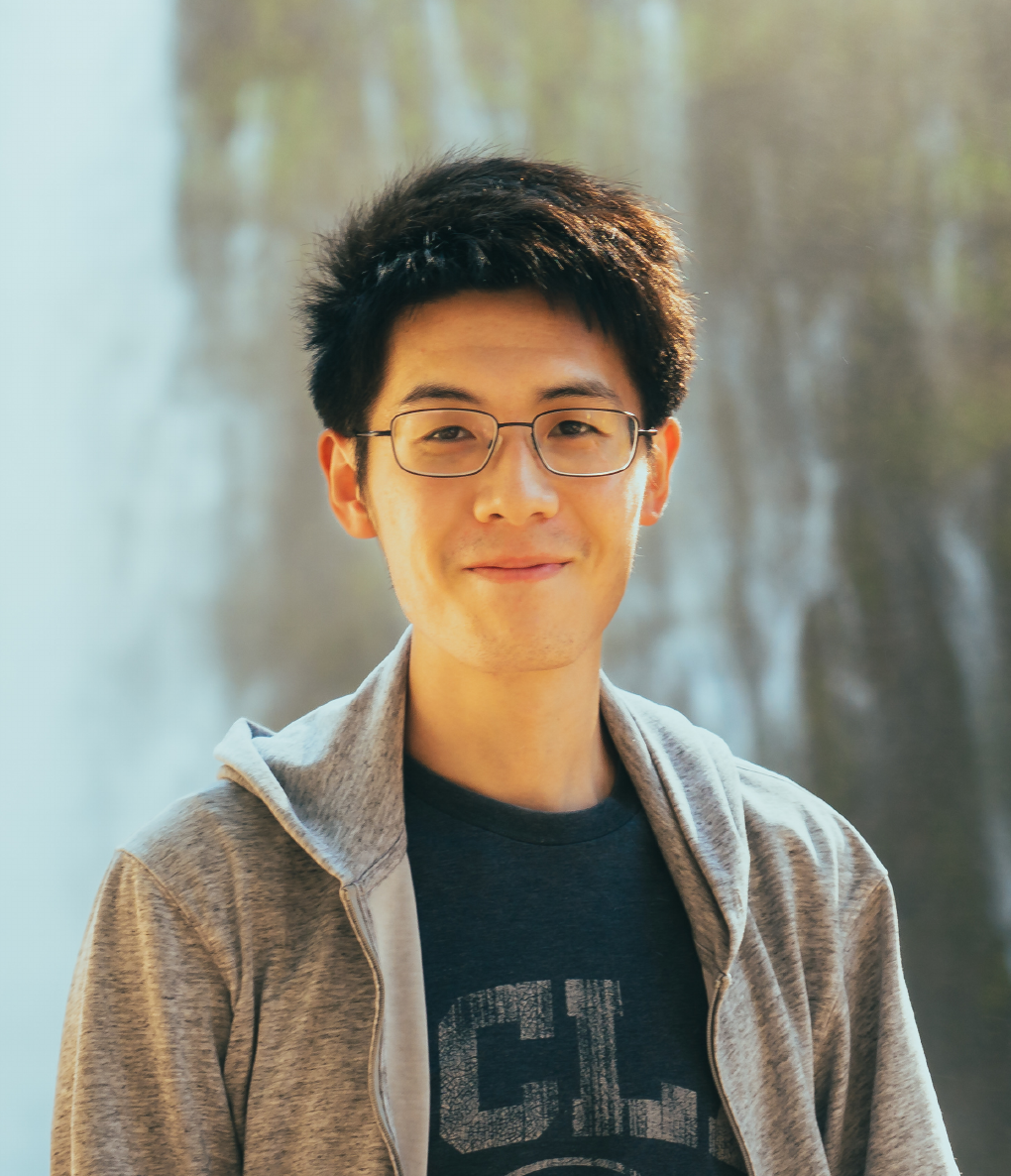}}]
{Licheng Guo}
received his B.S. degree in Electrical Engineering from Zhejiang University in 2018. Currently he is a 4th-year Ph.D student in the UCLA CS department. His research focus on co-optimizing HLS compilers (from C++ to RTL) and physical design tools (from RTL to hardware) to improve the circuit maximal frequency and reduce the compilation time.
\end{IEEEbiography}

\begin{IEEEbiography}[{\includegraphics[width=1in,height=1.25in,clip,keepaspectratio]{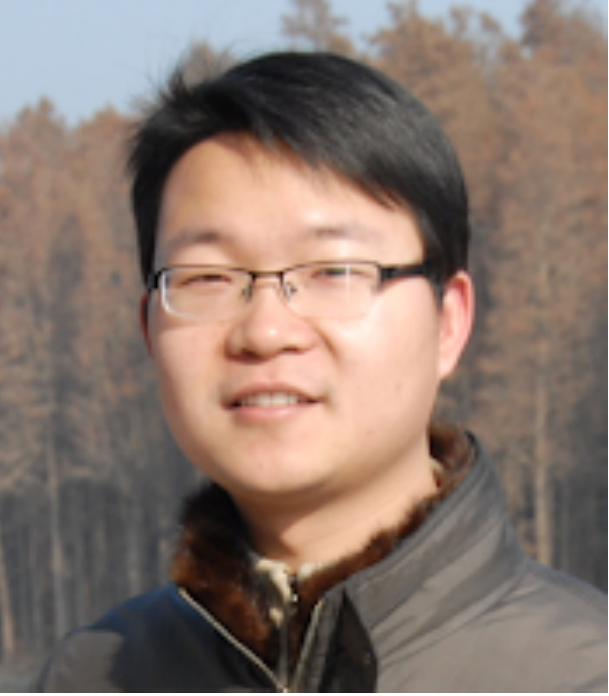}}]
{Zhenman Fang}
received his PhD degree in Computer Science from Fudan University, China in 2014. He did his postdoc at UCLA from 2014 to 2017, and worked as a Staff Software Engineer at Xilinx, San Jose, from 2017 to 2019. Currently, Zhenman is an Assistant Professor in School of Engineering Science, Simon Fraser University, Canada. His recent research focuses on customizable computing with specialized hardware acceleration, including emerging application characterization and acceleration, novel accelerator-rich and near-data computing architecture designs, and corresponding programming, runtime, and tool support. He is a member of the IEEE and ACM.
\end{IEEEbiography}

\begin{IEEEbiography}[{\includegraphics[width=1in,height=1.25in,clip,keepaspectratio]{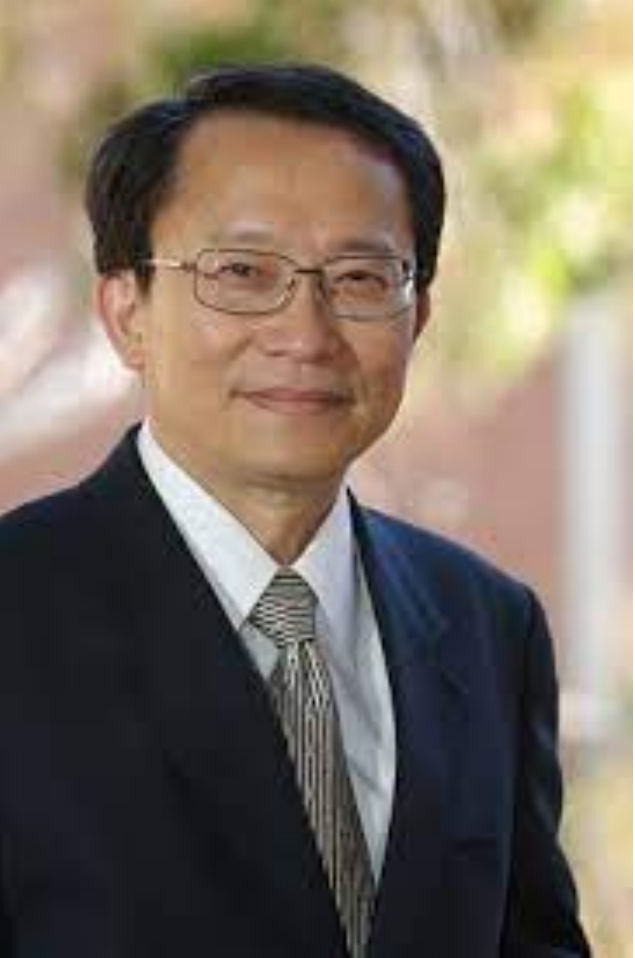}}]{Mau-Chung Frank Chang} 
is the Wintek Chair in Electrical Engineering and Distinguished Professor at UCLA. He pioneered the development of the world’s first multi-gigabit/sec data converters in heterojuction bipolar technologies; first mm-Wave Radio-on-Chip with Digitally Controlled on-chip Artificial Dielectric (DiCAD) for broadband frequency tuning and on-chip sensing/actuating with cautious feedback control for achieving self-diagnosis and self-healing capabilities. He realized the first CMOS frequency synthesizers up to terahertz spectra and demonstrated tri-color and 3-dimensional CMOS active imagers at the (sub)-mmWave spectra based on a time-encoded digital architecture. He is a Member of the US National Academy of Engineering, a Fellow of US National Academy of Inventors, an Academician of Academia Sinica of Taiwan, and a Lifetime Fellow of IEEE. He was honored with the IEEE David Sarnoff Award in 2006 for developing and commercializing GaAs HBT and BiFET power amplifiers, which dominated smartphones transmitter worldwide production throughout the past 2.5 decades. 

\end{IEEEbiography}

\begin{IEEEbiography}[{\includegraphics[width=1in,height=1.25in,clip,keepaspectratio]{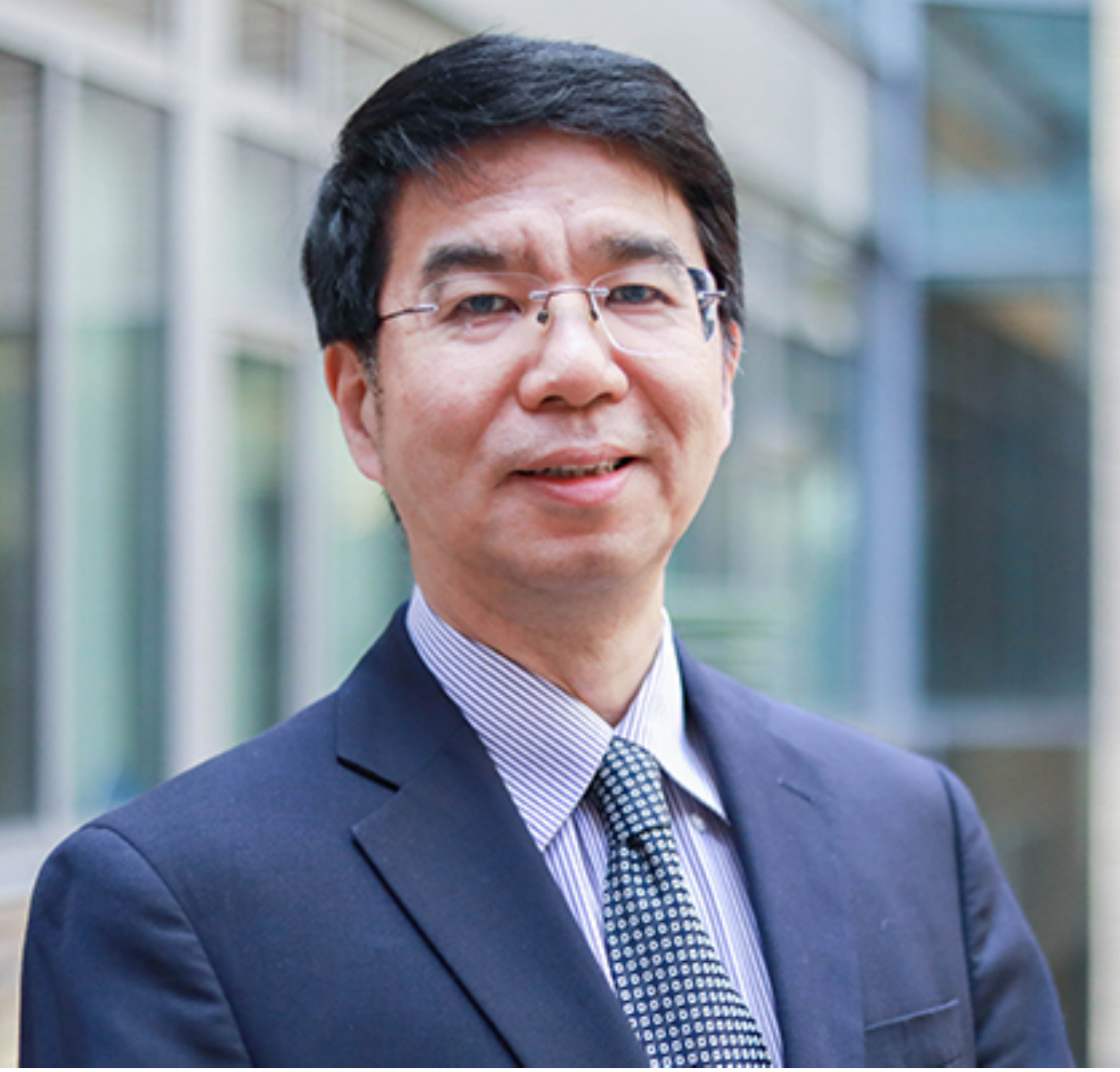}}]
{Jason Cong}
received his B.S. degree in computer science from Peking University in 1985, his M.S. and Ph. D. degrees in computer science from the University of Illinois at Urbana-Champaign in 1987 and 1990, respectively.  Currently, he is the Volgenau Chair for Engineering Excellence (and former department chair) at the UCLA Computer Science Department, with joint appointment from the Electrical Engineering Department. Dr. Cong’s research interests include novel architectures and compilation for customizable computing and quantum computing.  He has over 500 publications in these areas, including 16 best paper awards, three 10-Year Most Influential Paper Awards, and three papers in the FPGA and Reconfigurable Computing Hall of Fame.  He was elected to an IEEE Fellow in 2000, an ACM Fellow in 2008, a member of the National Academy of Engineering in 2017, and a Fellow of the National Academy of Inventors in 2020.
\end{IEEEbiography}